\documentclass[numberedappendix]{emulateapj}

\usepackage{natbib}
\usepackage{verbatim}
\renewcommand{\d}{{\rm d}}
\newcommand{\beq}{\begin{equation}}
\newcommand{\eeq}{\end{equation}}

\shorttitle{Where is the matter in Abell 2218?}
\shortauthors{El\'\i asd\'ottir et al.}

\begin{document}
\bibliographystyle{apj}

\title{Where is the matter in the Merging Cluster Abell 2218?
\thanks{Based on observations made with the Hubble Telescope's Advance Camera for Surveys, Chandra X-ray Observatory, the William Herschel Telescope and LRIS on the KECK telescope.} 
}



\author{\'Ard\'\i s El\'\i asd\'ottir\altaffilmark{2}, Marceau Limousin\altaffilmark{2}, Johan Richard\altaffilmark{3}, Jens Hjorth\altaffilmark{2}, Jean-Paul Kneib\altaffilmark{4},  Priya Natarajan\altaffilmark{5,6}, Kristian Pedersen\altaffilmark{2}, Eric Jullo\altaffilmark{7}, Danuta Paraficz\altaffilmark{2,8}}

\email{ardis@dark-cosmology.dk}




\altaffiltext{2}{Dark Cosmology Centre, Niels Bohr Institute, University of
Copenhagen, Juliane Maries Vej 30, DK-2100 Copenhagen \O, Denmark; 
(ardis, marceau, danutas, kp, jens)@dark-cosmology.dk.}
\altaffiltext{3}{Department of Astronomy, California Institute of  Technology, 105-24, Pasadena, CA91125; 5(johan,kneib)@astro.caltech.edu}
\altaffiltext{4}{OAMP, Laboratoire d'Astrophysique de Marseille UMR 6110 Traverse du Siphon 13012 Marseille, France; Jean-Paul.Kneib@oamp.fr}
\altaffiltext{5}{Astronomy Department, Yale University, P.O. Box 208101, New Haven, CT 06520-8101, USA, priya@astro.yale.edu}
\altaffiltext{6}{Department of Physics, Yale University, P.O. Box 208101, New Haven, CT 06520-8101; USA}
\altaffiltext{7}{European Southern Observatory, Alonso de Cordova 3107, Vitacura, Chile; ejullo@eso.org }
\altaffiltext{8}{Nordic Optical Telescope (NOT), Apartado 474, 38700 Santa Cruz de La Palma, Canary Islands, Spain}

\begin{abstract}
We present a parametric strong lensing model of the cluster Abell 2218 based on HST ACS data.  We constrain the lens model using 8 previously known multiply imaged systems, of which 7 have spectroscopically confirmed redshifts.   In addition, we propose five candidate multiply imaged systems and estimate their redshifts using our mass model.  
The model parameters are optimized in the source plane by a bayesian Monte Carlo Markov Chain as implemented in the the publicly available software Lenstool.  We find rms$_s=0\farcs12$ for the scatter  of the sources in the source plane, which translates into rms$_i=1\farcs49$ between the predicted and measured image positions in the image plane.
We find that the projected mass distribution of Abell 2218 is bimodal, which is supported by an analysis of the light distribution.  We also find evidence for two structures in velocity space, separated by $\sim 1000$~km~$s^{-1}$, corresponding to the two large scale dark matter clumps.  We find that the lensing constraints can not be well reproduced using only dark matter halos associated with the cluster galaxies, but that the dark matter is required to be smoothly distributed in large scale halos.   At $100\arcsec$ ($291$~kpc) the enclosed projected mass is $3.8\times10^{14}$~M$_\sun$.  At that radius, the large scale halos contribute $\sim85\%$ of the mass, the brightest central galaxy (BCG) contributes  $\sim9\%$ while the remaining $\sim6\%$ come from the other cluster galaxies.  We find that the model is not very sensitive to the fainter (and therefore by assumption less massive) galaxy sized halos, unless they locally perturb a given multiply imaged system.  Therefore, dark galaxy sized substructure can be reliably constrained only if it locally perturbs one of the systems.  The massive BCG and galaxies which locally perturb a multiply imaged system are reliably detected in the mass reconstruction.  In an appendix we give a self-contained description of the parametric profile we use, the dual pseudo isothermal elliptical mass distribution (dPIE).  This profile is a two component pseudo isothermal mass distribution (PIEMD) with both a core radii and a scale radii.
\end{abstract}

\keywords{ dark matter --- galaxies: clusters: individual (Abell 2218) --- gravitational lensing}

\section{Introduction}
Dark matter dominates over baryonic matter in the universe, but its nature is not known.  The study of the inner parts of dark matter halos can give insight into the nature of the dark matter, as the steepness of the profile is correlated with the interaction between the dark matter itself and with the baryonic matter.  According to $\Lambda$CDM simulations, the mass distribution of galaxy clusters should be dominated by their dark matter halos.  Gravitational lensing, which is sensitive to the total matter distribution, visible or dark, is ideal for studying the mass distribution of clusters.  Strong lensing features, consisting of multiply imaged and strongly distorted background sources, provide constraints on the inner parts of the cluster, while weak lensing features, consisting of weakly distorted singly imaged background sources, provide constraints on the outer slope of the surface mass profile \citep[see e.g.,][]{smail1995a,smail1995b,kneib1996, smail1997,abdelsalam1998,natarajan2002b, bradac2006,gavazzi2007,limousin2007}.

Lensing can therefore provide unique information about the total mass distribution of clusters, from the inner to the outer parts.  In addition, lensing can in principle be used to deduce various cosmological parameters (e.g. $H_0$, $\Omega_\Lambda$, $\Omega_m$).   This has already been extensively applied to lensing on galaxy scales \citep[see e.g.,][]{schechter1997,koopmans2003} and to a smaller degree for lensing on cluster scales \citep[see e.g.,][]{soucail2004, meneghetti2005}, where the accuracy of the mass map is a limiting factor.  The accuracy of the mass map is strongly dependent on the number of multiply imaged systems used to constrain it. Therefore, to construct a robust model of the dark matter distribution, accurate enough for cosmography and for using the cluster as a gravitational telescope, it is important to include as many spectroscopically confirmed multiply imaged systems as possible \citep[see e.g.,][]{ellis2001}.

Abell 2218 is one of the richest clusters in the Abell galaxy cluster catalog \citep{abell1958, abell1989} and has been successfully exploited as a gravitational lens.    
A parametric lens model has previously been constructed by \citet{kneib1995,kneib1996} (using 1 and 2 spectroscopically confirmed systems respectively) and by \citet{natarajan2002b, natarajan2007} (using 4 and 5 spectroscopically confirmed systems) building on the model of \citet{kneib1996} and including weak lensing constraints from HST WFPC2 data.  A non-parametric model was constructed by \citet{abdelsalam1998} using three spectroscopically confirmed multiply imaged systems.  In all of these models, a bimodal mass distribution was required to explain the image configurations (i.e. the models include two large scale dark matter clumps), but the number of constraints were not sufficient to accurately constrain the second large scale dark matter clump.  Abell 2218 has also been used as a gravitational telescope, with  \citet{ellis2001} discovering a source at $z=5.6$ and \citet{kneib2004} discovering an even more distant source at $z\sim6.7$, later confirmed by \citet{egami2005} using Spitzer data.  \citet{soucail2004} estimated cosmological parameters based on a lensing model of Abell 2218 using 4 multiply imaged systems.  The latest published lens model of Abell 2218 is by \citet{smith2005}, who incorporated four multiply imaged systems and weak lensing constraints, using the WFPC2 data.  Although the number of constraints has increased from the initial models, all previous models have assumed that the location of the second dark matter clump coincided with the brightest galaxy in the South East, due to a lack of constraints in its vicinity.

The motivation for revisiting the modeling of Abell 2218 comes from the new ACS data which have not been used before for modeling this cluster and are superior in both resolution, sensitivity and field of view to the previous WFPC2 data set.  These new high quality data allow us to identify several subcomponents in previously known multiple images, thus adding more constraints, and in one case, two more multiple images of a known system.
In addition, we have measured a spectroscopic redshift for an arc around the second dark matter clump, which our model predicts to be singly imaged.
We also have several new candidate multiply imaged systems, which we add as constraints and estimate their redshift with the model.  In total we have 7 multiply imaged systems with measured spectroscopic redshift and 6 multiply imaged systems without spectroscopic data (of which 5 are new candidate systems).  Finally, the lensing code Lenstool, has undergone significant improvements from previous models, with the incorporation of a Monte Carlo Markov Chain (MCMC), which enables us to not only find the best model in the lowest $\chi^2$ sense, but the most likely model as measured by its Evidence \citep{jullo2007}.  The MCMC also allows for a reliable estimate of the uncertainties in the derived model parameters. 

The paper is organized as follows:  In Section~\ref{sec:data}, we give an overview of the data used in this paper.  We compile a list of all currently known and new multiply imaged systems in Abell 2218 and discuss the reliability of the redshift estimate of each one in Section~\ref{sec:systems}.  The methodology of the strong lensing modeling is described in Section~\ref{sec:modeling}.    In Section~\ref{sec:analysis} we present the results of our lensing analysis, and compare them to previous models.  In Section~\ref{sec:degeneracy} we discuss degeneracies in the modeling.  In Section~\ref{sec:reliability} we address how reliable our model is, and discuss the smoothness of the dark matter distribution.
 In Section~\ref{sec:bimodal} we interpret the bimodality of our model, along with X-ray measurements and an analysis of the distribution of cluster members in velocity space.  We summarize our main conclusions in Section~\ref{sec:conclusions}.   Throughout the paper, we adopt a flat $\Lambda$-dominated Universe with $\Omega_\Lambda=0.7$, $\Omega_m=0.3$ and $H_{0}=70\ \mathrm{km\,s}^{-1}\,\mathrm{Mpc}^{-1}$.  Following \citet{smith2005} we place the cluster at $z=0.171$.  At this redshift $1\arcsec$ corresponds to $2.91$~kpc in the given cosmology.  

\section{Data}
\label{sec:data}
We use data from several different sources for our lens modeling and analysis.  The basis of our modeling is Advanced Camera for Surveys (ACS) data from the Hubble Space Telescope, which allows us to identify and accurately locate multiply imaged systems.  Our cluster member catalog is also selected using the ACS data and the magnitude of each cluster member is given in the Ingrid K-band of the William Herschel Telescope.  In addition, we have obtained a spectroscopic redshift of a system using the Keck telescope and archival Chandra X-ray Observatory data have been used to produce an X-ray map of Abell 2218. 

\subsection{The galaxy catalog}
\label{sec:galaxycat}
Cluster members were selected based on the ACS data using the characteristic cluster red-sequences (V-Z) and (I-Z) in two color-magnitude diagrams and selecting objects lying within the red-sequences of 
\begin{figure}
\includegraphics[angle=270, scale=0.4]{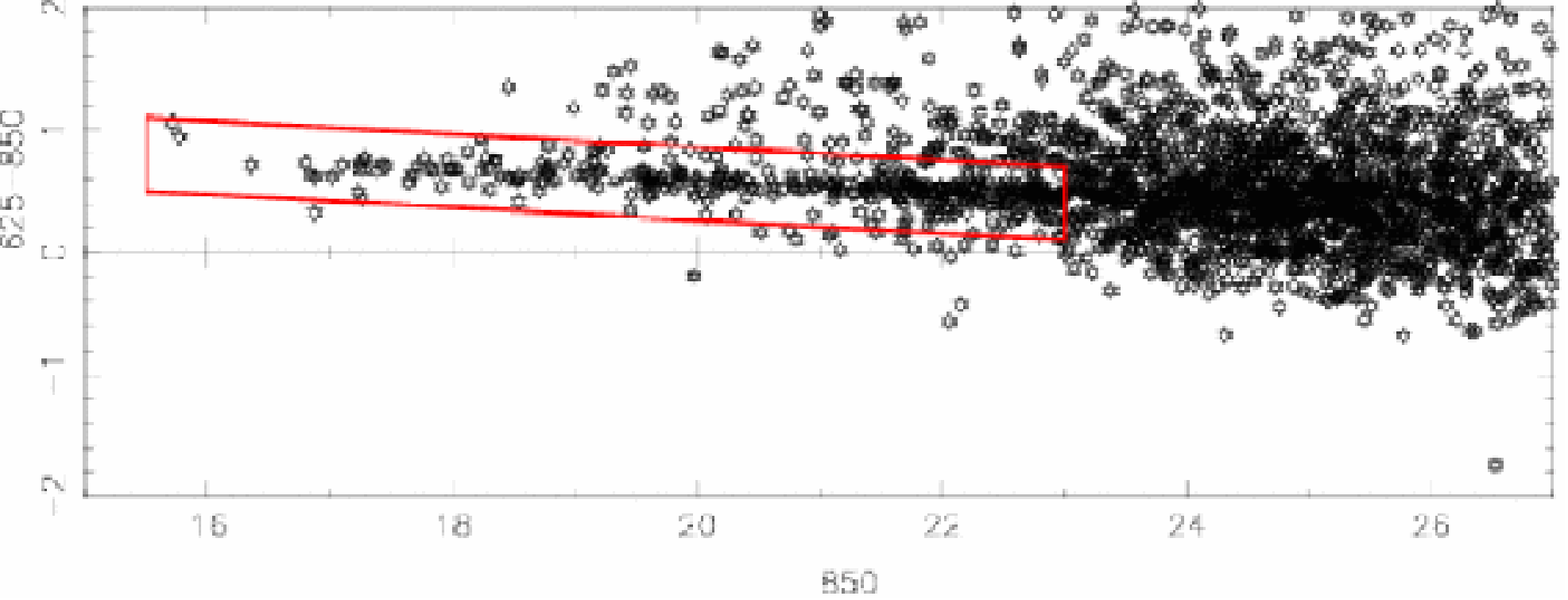}\\
\includegraphics[angle=270, scale=0.4]{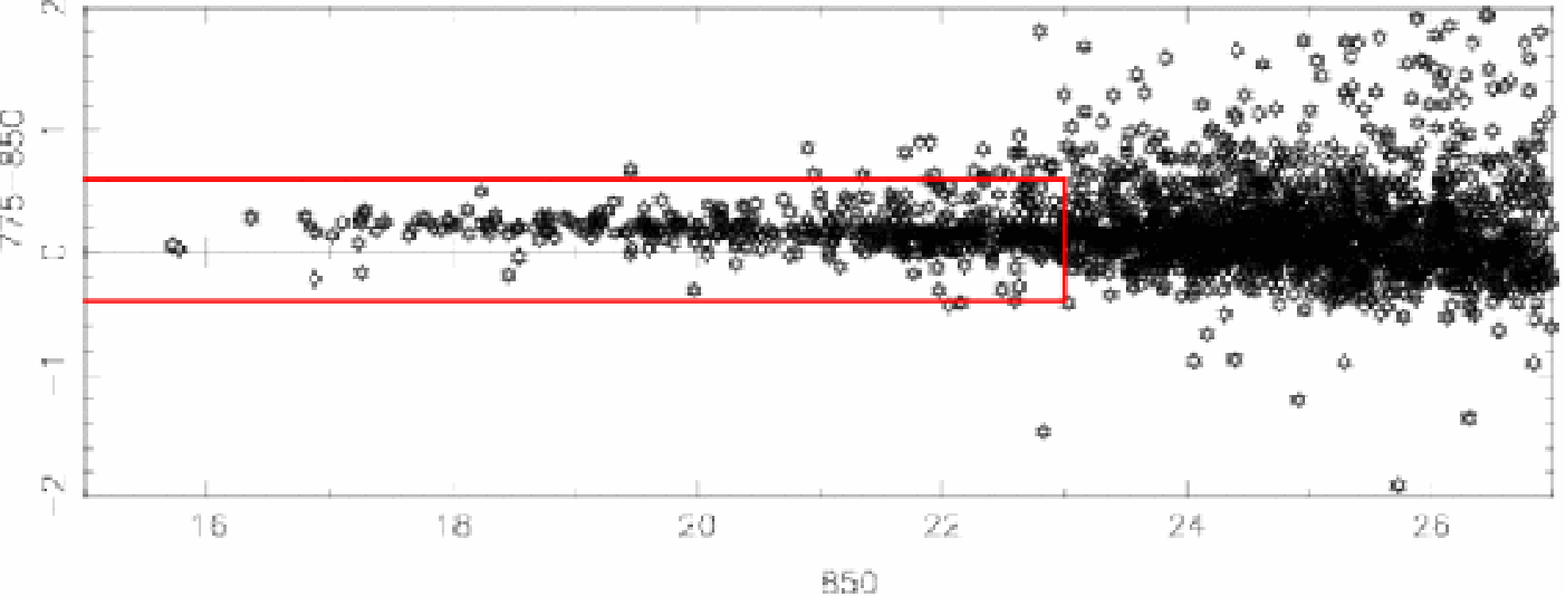}
\caption{Color-magnitude diagrams, (V-Z) and (I-Z), used for selecting the cluster members (see Section~\ref{sec:galaxycat}).  The red quadrilaterals define the red sequence and galaxies lying within are considered cluster members.   \label{fig:redsequence}}
\end{figure}
\begin{figure*}
\epsscale{1.0}
\plotone{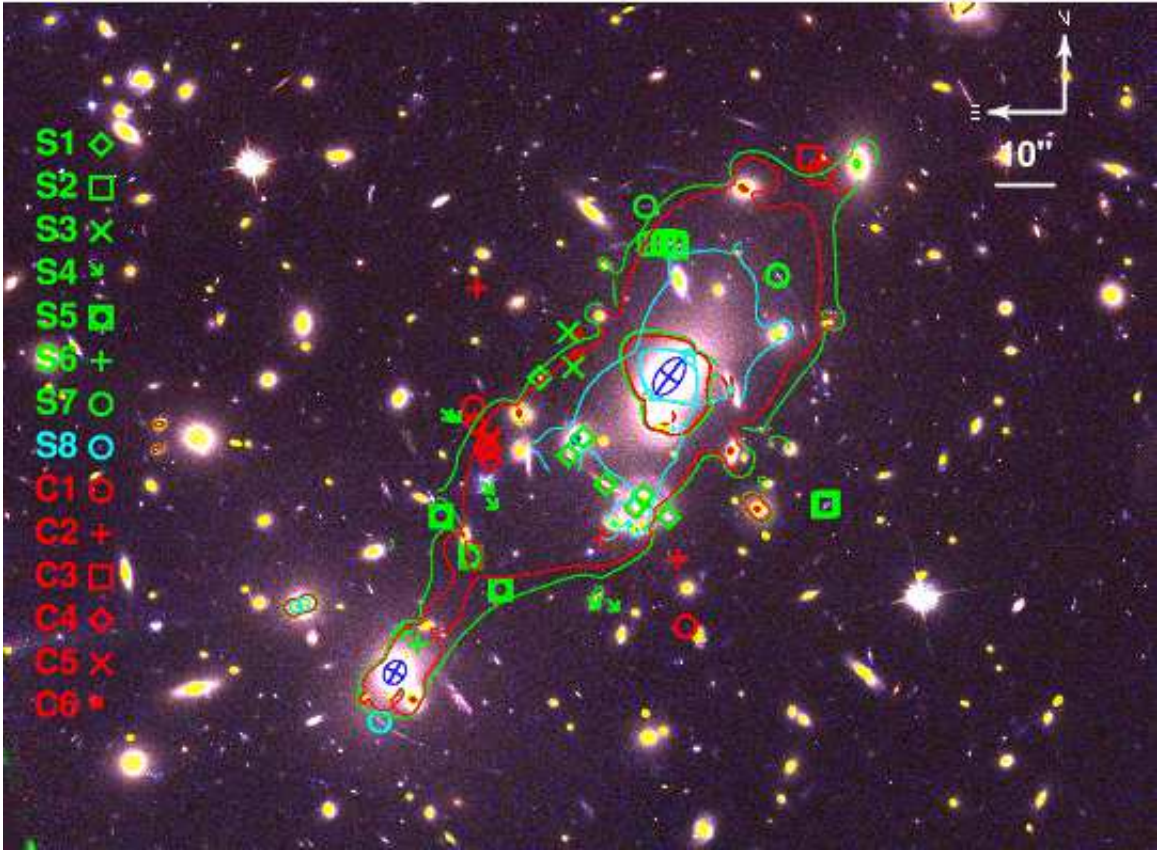}
\caption{A color image of Abell 2218 based on ACS data (F775W, F625W and F475W filters in the red, green and blue channel respectively).  The cluster galaxies are marked in yellow (modeled using scaling relations) or blue (individually fitted).  The multiple images are labeled in green for spectroscopically confirmed systems and red for candidate systems.  The arc for which we have obtained spectroscopic redshift, S8, is labelled in cyan.  Also shown are the critical lines corresponding to $z=0.702$ (cyan), $z=2.515$ (red) and $z=6.7$ (green).  \label{fig:overview}}
\end{figure*}
both diagrams (see Figure~\ref{fig:redsequence}).  The ACS data were used to select the galaxy members due to their better photometric accuracy compared to the K-band data.  Nevertheless, the catalog is given in the K band since it is more representative of the galaxy population in elliptical galaxies.  This selection gave  $203$ cluster galaxies down to $K=19.6$.  However, we rejected six galaxies from the catalog leaving $197$ cluster members, of which four have measured redshifts showing them to be background galaxies.

\subsection{X-ray data}
The X-ray data were taken with the Advanced CCD Imaging Spectrometer (ACIS) on NASA's Chandra X-ray Observatory.  The ACIS is an array of 10 CCD's, capable of simultaneously imaging and measuring the energy of the incoming X-rays.  We combined three images taken at different epochs to construct our final X-ray map of Abell 2218.  Two of the images were taken in October 1999 and one in August 2001, with 5960~s, 11560~s and 49240~s exposure times respectively.  The data reduction was performed using standard pipelines from the Chandra Interactive Analysis of Observations (CIAO) software.  The X-ray data will be used for comparison with the lensing data in Section~\ref{sec:bimodal}.

\section{Multiply imaged systems}
\label{sec:systems}
Multiply imaged systems are the basis of our analysis, and are particularly useful as constraints if their redshifts are accurately determined.  In this section we list the multiple images, with or without spectroscopically confirmed redshifts, used as constraints in our lens modeling.  We propose several new candidate systems found in the ACS data and present a spectroscopic redshift measurement for an arc which lies in the region of the second dark matter clump.  All systems are listed in Table \ref{tab:systems} and shown on Figure~\ref{fig:overview}.

\begin{deluxetable}{lllll} 
\tablecolumns{5} 
\tablewidth{0pc} 
\tablecaption{Lensed systems} 
\tablehead{ 
\colhead{System id}    & \colhead{R.A (J2000)} & \colhead{Dec. (J2000)} 
&   \colhead{$z$\tablenotemark{$\dagger$}}   & \colhead{Comments}  }
\startdata 
S1.a & 248.96976  & 66.212219 & 0.702 &  (1) \\
S1.b & 248.96646  & 66.208627 & 0.702 &   \\
S1.c  & 248.96478  & 66.209439 & 0.702 &   \\
S1.d  & 248.96218  & 66.207296 & 0.702 &   \\
S1.e & 248.95876  & 66.206237 & 0.702 &   \\
S1.f & 248.95790  & 66.206762 & 0.702 &   \\
S1.g & 248.95500  & 66.205771 & 0.702 &   \\
\hline
S2.1.a & 248.95540 & 66.218525 & 2.515 & (2)\\
S2.1.b & 248.95451 & 66.218469 & 2.515 & \\
S2.1.c & 248.93684 & 66.206345 & 2.515 & \\
S2.2.a & 248.95562 & 66.218536 & 2.515 & \\
S2.2.b & 248.95426 & 66.218458 & 2.515 & \\
S2.2.c & 248.93681 & 66.206424 & 2.515 & \\
S2.3.a & 248.95642 & 66.218428 & 2.515 & \\
S2.3.b & 248.95441 & 66.218347 & 2.515 & \\
S2.3.c & 248.93723 & 66.206355 & 2.515 & \\
S2.4.a & 248.95405 & 66.218258 & 2.515 & \\
S2.4.b & 248.95715 & 66.218375 & 2.515 & \\
S2.4.c & 248.93728 & 66.206269 & 2.515 & \\
\hline
S3.a & 248.96646 & 66.214324 & 5.576 & \\
S3.b & 248.96591 & 66.212701 & 5.576 & \\
\hline
S4.1.a & 248.97894 & 66.210224 & 2.515 & \\
S4.1.b & 248.97519 & 66.206897 & 2.515 & \\
S4.1.c & 248.96075 & 66.201462 & 2.515 & \\
S4.2.a & 248.97981 & 66.210279 & 2.515 & \\
S4.2.b & 248.97463 & 66.206179 & 2.515 & \\
S4.2.c & 248.96296 & 66.201575 & 2.515 & \\
\hline
S5.a & 248.981043 & 66.205816 & 2.515 & (3)\\
S5.b & 248.977680 & 66.203939 & 2.515 & \\
S5.c & 248.974235 & 66.202425 & 2.515 & \\
\hline
S6.a  & 248.9855   & 66.200483 & 1.034 & (4)\\
S6.b  & 248.98371 & 66.199906 & 1.034 & \\
\hline
S7.a & 248.94254 & 66.216974 & 1.03 & (5)\\
S7.b & 248.94244 & 66.216909 & 1.03 & \\
S7.c & 248.95763 & 66.220132 & 1.03 & \\
\hline
S8.a & 248.98805& 66.19633  & 2.74& \\ 
\hline
C1.a & 248.97735 & 66.210845 & $\sim$6.7 & (6) \\
C1.b & 248.97585 & 66.208989 & $\sim$6.7 & \\
C1.c & 248.95303 & 66.200703 & $\sim$6.7 & \\
\hline
C2.a  & 248.95398 & 66.203803 &  [$2.6\pm0.1$]& \\
C2.b  & 248.96247 & 66.204748 &   [$2.6\pm0.1$]& \\
C2.c  & 248.97692 & 66.216365 &  [$2.6\pm0.1$]& \\
\hline
C3.a &  248.93879  & 66.22238   & [$2.8\pm0.6$]& \\
C3.b &    248.93749  & 66.221662   &[$2.8\pm0.6$]&\\
\hline
C4.a  &  248.97708   & 66.210108   &  [$2.2\pm0.2$] & \\
C4.b &  248.97542   & 66.208369   &  [$2.2\pm0.2$] & \\
\hline
C5.a  & 248.97543 & 66.209244 &   [$2.3\pm0.8$]& \\
C5.b  & 248.97556 & 66.209383 &  [$2.3\pm0.8$]& \\
\hline
C6.a & 248.96481 & 66.213039  &  [$2.6\pm0.3$] & \\
C6.b & 248.96505 & 66.214017 &   [$2.6\pm0.3$]& \\
\enddata 
\tablecomments{The multiply imaged systems used to constrain the model.  The systems and their properties, along with refences to the papers reporting their redshifts, are given in Section~\ref{sec:systems}.  Systems C2-C6 are new candidate systems reported in this paper.  Systems with a spectroscopic redshift measurement are denoted by an S in their id, while those without spectroscopic redshifts  are denoted by a C.  Roman letters (a,b,c,...) denote the different images of each set of multiple images.  If several components are identified in system, each component is sublabelled, e.g. 2.1 and 2.2 denote two different components of the same multiple imaged system.  (1) Two pairs of images are caused by splitting by a local cluster galaxy. (2) Also known in the literature as \#384 and \#486. (3) The redshift is measured for an associated singly imaged galaxy  (4) Also known in the literature as \#289   (5) Also known as \#444 and H6.  The redshift estimate is based on one feature.  (6) Redshift estimate based on photometric redshift, lensing and a tentative Ly$\alpha$ break.} 
\tablenotetext{$\dagger$}{Lens model redshift predictions are reported in square brackets.}
\label{tab:systems} 
\end{deluxetable}

\subsection{Previously known systems}
\label{sec:prev_known}
The first system, system S1, is at a redshift of $z=0.702$ as measured by \citet{leborgne1992} (system 359 in their catalog).  Thanks to the new ACS observations, it has seven identified images (previously it had five identified images \citep{kneib1996}), of which two pairs are located close to individual galaxies $\#1028$ and $\#993$ in the galaxy catalog) which affect the lensing signal.

System S2 is a star-forming galaxy at a redshift of $z=2.515$ \citep{ebbels1996}.  It consists of a fold and a counter image (listed as images $\#384$ and $\#468$ in \citet{ebbels1996}).  We separate the fold image into two images and identify four components, giving us four sets of triple images from this system (referred to as S2.1, S2.2, S2.3 and S2.4). 

System S3 is a faint star-forming doubly imaged system at a high redshift $z=5.576$ discovered by \citet{ellis2001}.  

System S4 is a triply imaged submillimeter source at $z=2.515$ reported by \citet{kneib2004b}.  In each image they identify three components labelled $\alpha, \gamma, \beta$.  We identify all three images in the ACS data.  However, we only detect the $\alpha$ and $\beta$ components.  We include both as constraints on the model (referring to them as S4.1 and S4.2 respectively).  Image S4.1 seems to consist of further three subcomponents.  As we do not clearly distinguish them in all three images, we do not add them as separate constraints.

The fifth system, S5, is a triply imaged system, believed to be the triply imaged outskirts of an associated singly imaged galaxy.  It is believed that the galaxy partially crosses the caustic, with the central part being singly imaged and the outskirts being triply imaged.  \citet{kneib2004} measured a redshift of $z=2.515$ for the galaxy (labelled as $\#273$ in their notation) but an independent spectroscopic redshift has not been determined for the faint triply imaged component.  Although it is likely (and consistent with the models) that they belong to the same background source, it is not certain, making the redshift determination for S5 less certain.

The sixth system, S6, consists of an arc in two parts at $z=1.034$ \citep[][called 289 in their notation]{kneib1996, swinbank2003}.  

The seventh system, S7, first reported by \citet{kneib1996}, consists of a merging image and a counter image (also known as \#444 and H6 respectively).  \citet{ebbels1998} measured a spectroscopic redshift at $z=1.03$ for the merging image
but this is listed as a tentative identification (as the only identified feature in the spectra was the Fe II doublet at $\lambda\lambda2587$, $2600$).  

The final previously known multiply imaged system, C1, is a high redshift ($z\sim7$) triply imaged system \citep{kneib2004}. \citet{egami2005} further constrained the redshift of the system to lie in the range $6.6-6.8$.  For the purposes of the modeling we will assign $z=6.7$ to this system, noting that changing the redshift within the range $6.6-6.8$ does not affect the model.

\subsection{New candidate systems}
We propose 5 new candidate systems which we have identified in the new ACS data.  We do not have spectroscopic redshift determination for these systems, but estimate their redshifts using the model prediction.  

\begin{figure}
\epsscale{.35}
\plotone{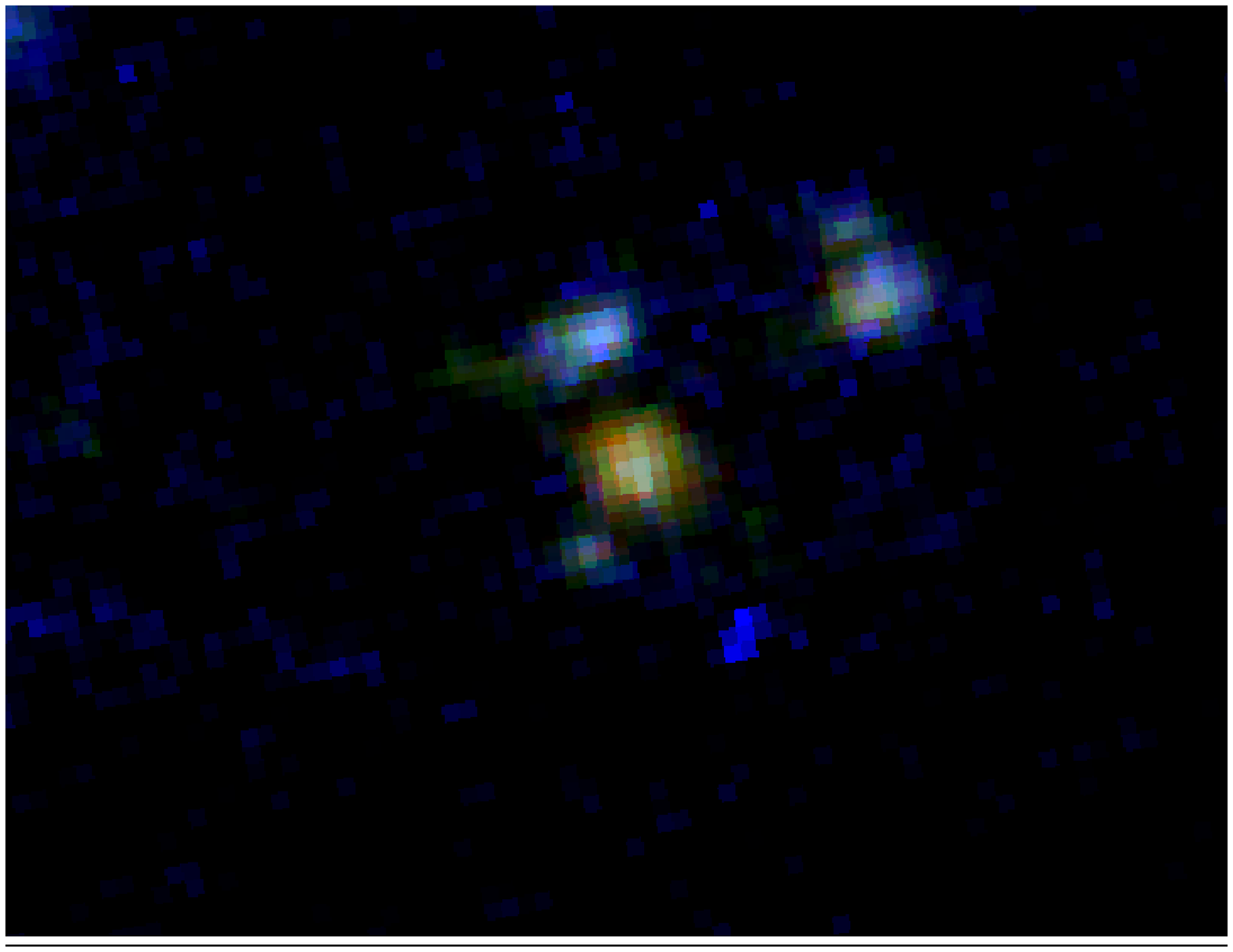}
\plotone{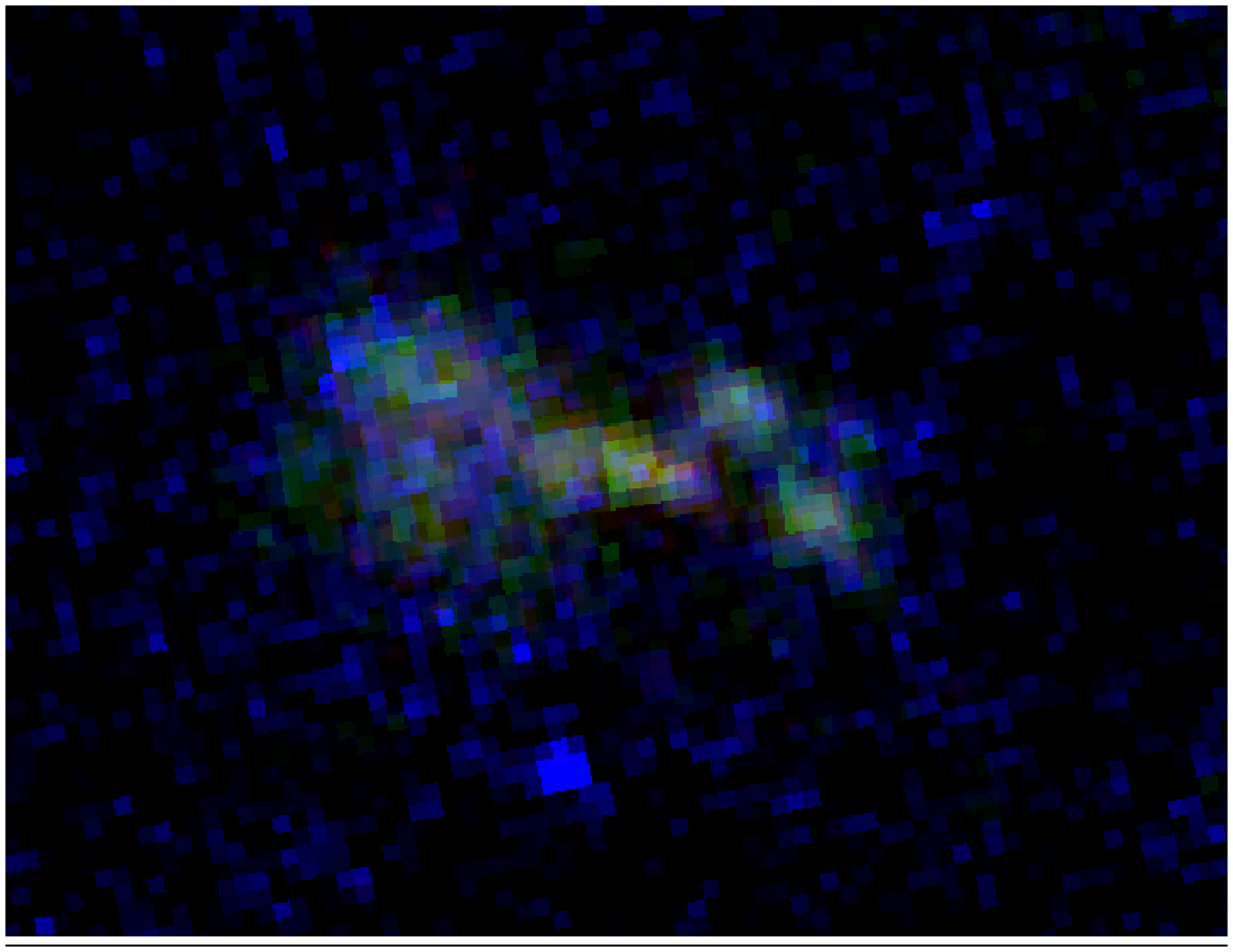}
\plotone{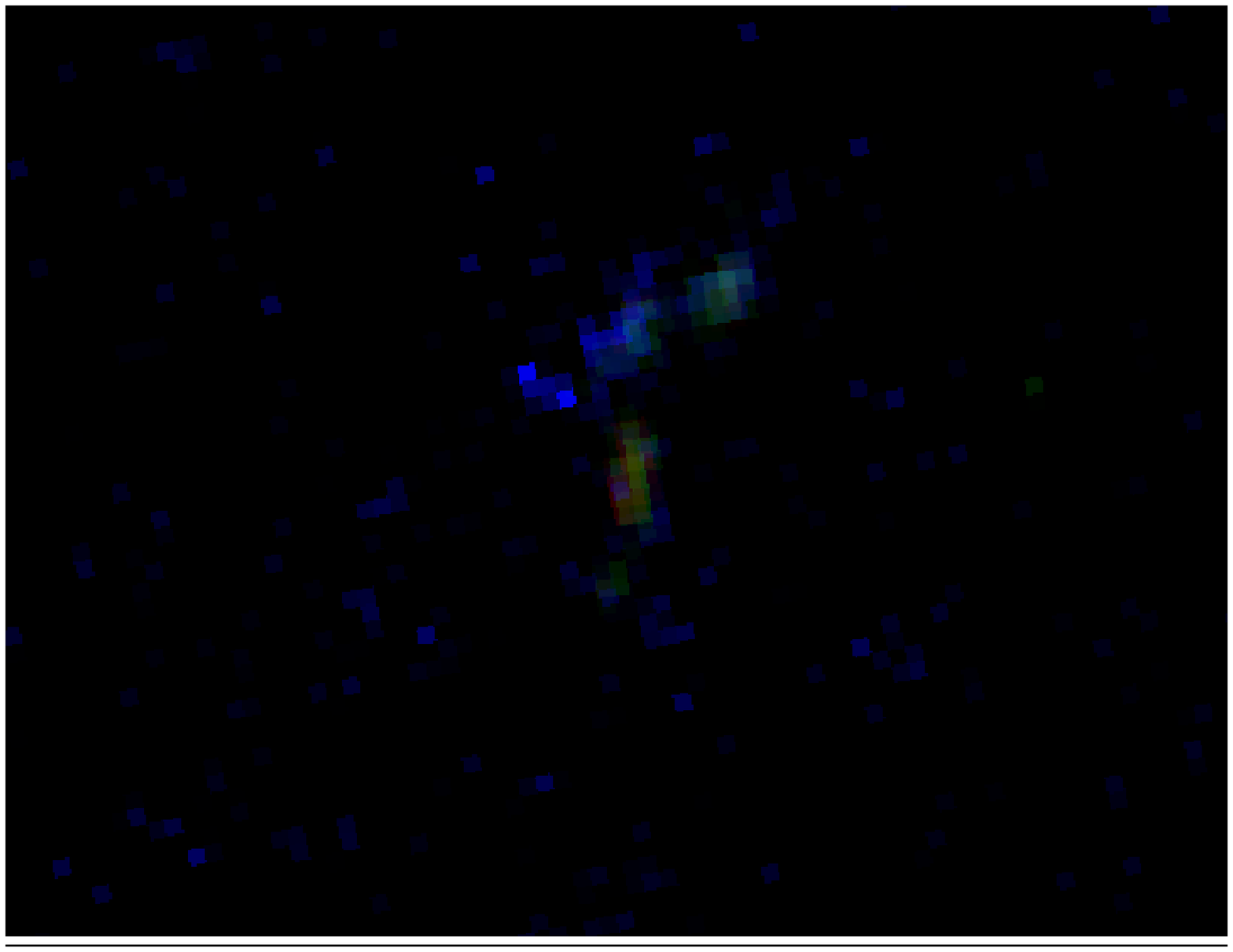}
\caption{The new candidate system, C2, identified in the ACS images.  The three panels show the three images of C2, marked in Figure~\ref{fig:overview}.  The colors are the same as in Figure~\ref{fig:overview}.  The system is characterized by a central 'yellow' spot, flanked by blue spots.  \label{fig:newsys}}
\end{figure}
The first candidate is a triply imaged system, which we call C2.  In Figure~\ref{fig:newsys} we show color stamps of the three images, and the locations are given in Table~\ref{tab:systems}.  The morphology of the three images, suggests that it is of the same background source, with a 'yellow' spot in the middle, flanked by a fainter blue on the sides.  

\begin{figure}
\epsscale{.35}
\plotone{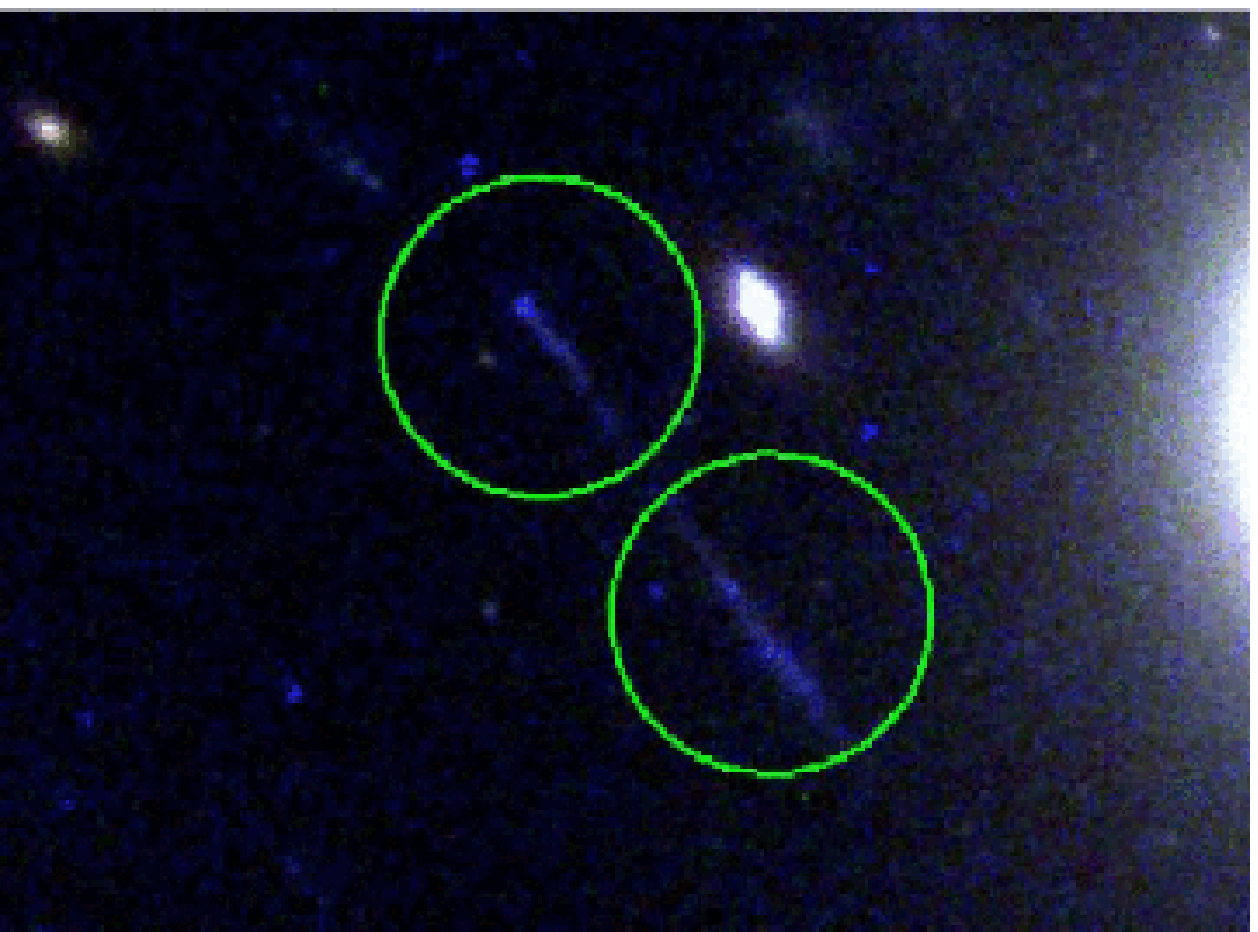}
\plotone{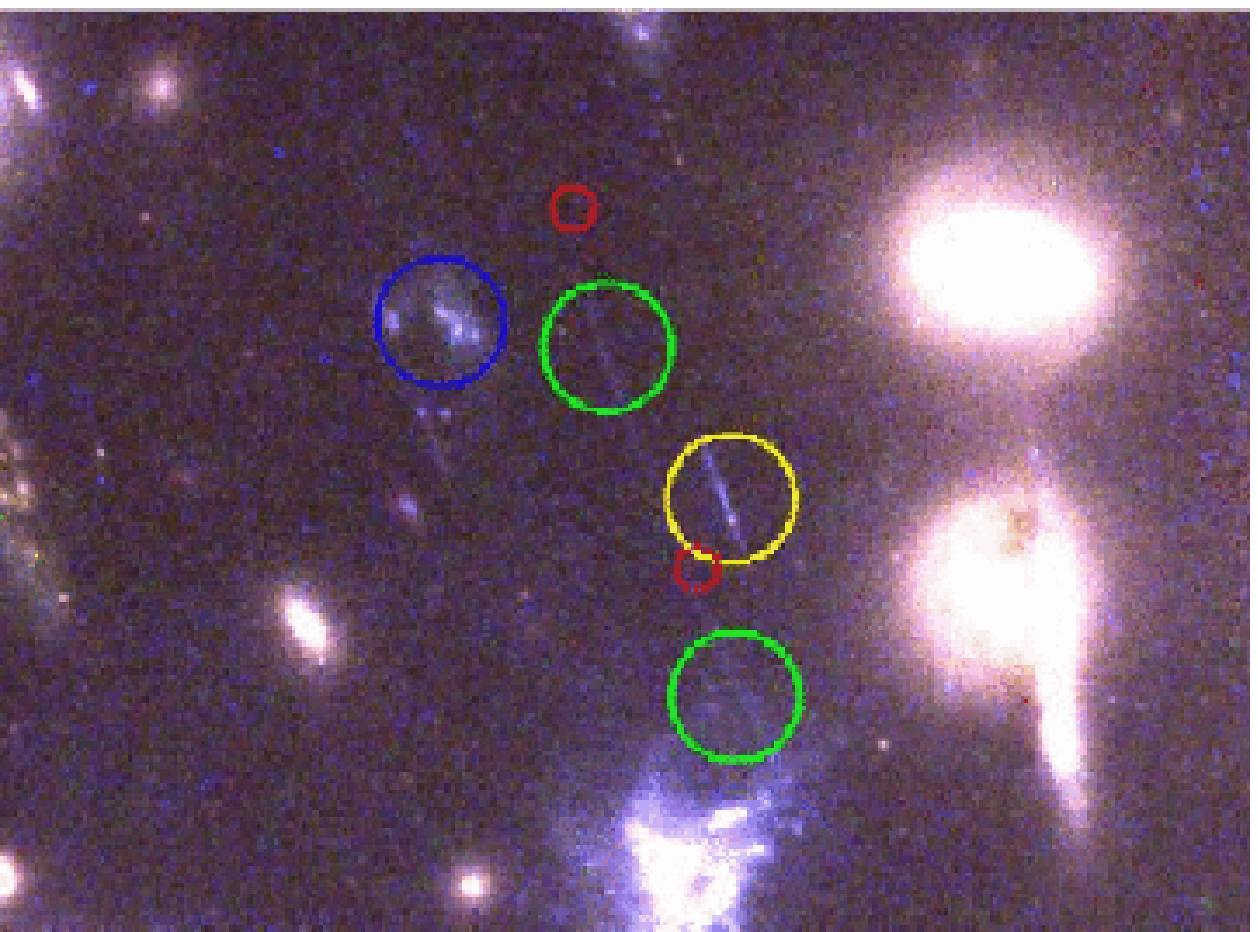}
\plotone{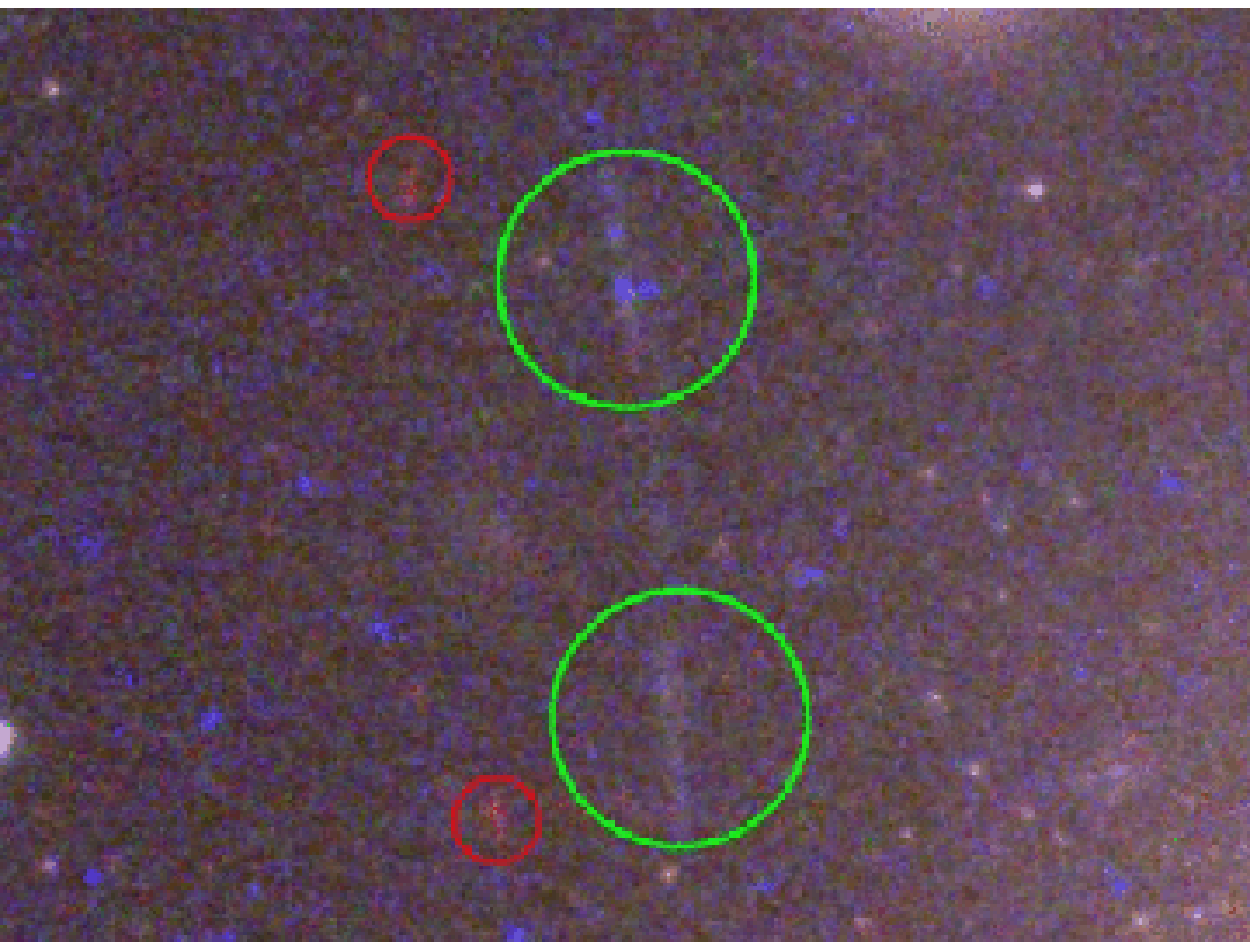}
\caption{The new candidate arcs identified in the ACS images.  {\it Left panel:}  The candidate arc C3 (green circles).  {\it Middle panel:}   The candidate systems C4 (green circles) and C5 (yellow circle).  The red circles are the locations of two of the images of system C1 (at $z=6.7$), while the blue circle shows the location of images belonging to system S4 (components a and b).   {\it Right panel:}  The candidate system C6 (green circles).  The high redshift system S3 ($z=5.6$) is marked with red circles.  \label{fig:newarcs}}
\end{figure}
We also identify a new faint arc in the north-west part of the cluster, which we call C3 (see Figure~\ref{fig:newarcs}).  
In the vicinity of system 6, we further identify two arclike images, which we label C4 and C5.  The first, C4, is a very faint blue double arc, while in the second, C5, two bright spots of a merging arc can be seen.    
Finally, a pair of blue extended images, C6, is found in the vicinity of system 3.
These candidate systems  are shown in a color image in Figure~\ref{fig:newarcs} and are listed in Table~\ref{tab:systems} along with their redshift estimates (see Section~\ref{sec:redshifts} for details).  The positions of these images are used to constrain the final model, but their redshift is kept free in the modeling. 

\subsection{Spectroscopic redshift for an arc}
\begin{figure}
\epsscale{0.8}
\plotone{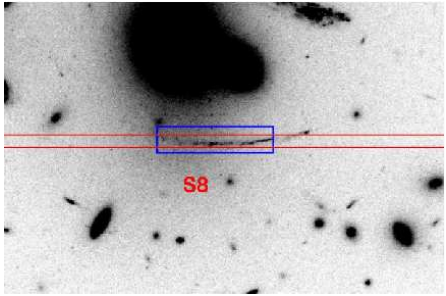}
\caption{The alignment of the slit for the spectra of S8.  The box shows the area used to extract the spectrum (see Figure~\ref{fig:S8_spectrum}).\label{fig:a2218_S8_slit}}
\end{figure}

We have used the Low Resolution Imager and Spectrograph (LRIS, \citet{oke1995}) at Keck to measure the spectroscopic redshift of an arc, S8, to the south-east of the second dark matter clump (see Figure~\ref{fig:overview}).  Two exposures of 1800 seconds were obtained on June 29th 2007 with a $175\arcsec\times1\arcsec$
 long slit placed along the
brighter components of this arc (Figure~\ref{fig:a2218_S8_slit}).
A $600$~l~mm$^{-1}$ grism blazed at $4000$~\AA~and a $400$~l~mm$^{-1}$
grating blazed at $8500$~\AA~were used in the blue and red
channels of the instrument, both lightpaths being separated by a dichroic at $5600$~\AA. The corresponding dispersions are $0.6$/$1.85$~\AA~and
resolutions  are $4.0$/$6.5$~\AA~in the blue/red channel, respectively.

\begin{figure}
\includegraphics[angle=270, scale=0.3]{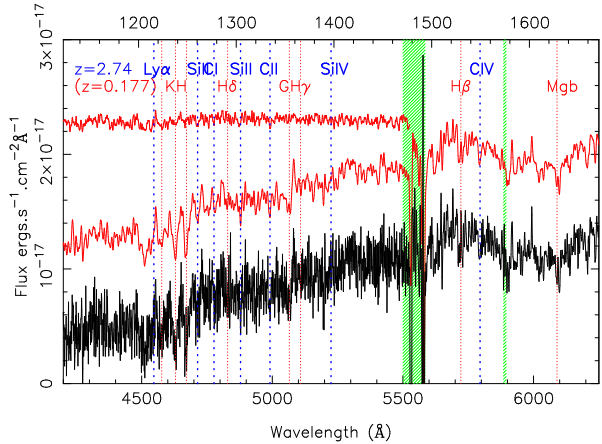}
\caption{The spectrum of system S8.  The plot shows lines for the arc at $z=2.74$ (blue) and the nearby galaxy (\#617) at $z=0.177$ (red).  The spectrum shows Ly$\alpha$ absorption and emission and a number of absorption lines in the UV which are marked in the figure.  At this redshift, $z=2.74$, the model predicts S8 to be singly imaged.  \label{fig:S8_spectrum}}
\end{figure}
The resulting spectrum (see Figure~\ref{fig:S8_spectrum}) is dominated by the light coming
from the very bright neighbor cluster member at $z=0.177$, yet shows
Lyman-$\alpha$ in emission and UV absorption features of SiII, CI, CII and
CIV from the arc, giving a redshift $z=2.74$. These features are not compatible with any line at the redshift of the lens galaxy, giving a
redshift class of 2 for this spectrum, with $75\%$ probability of being correct, following the classification of \citet{lefevre1995}.  Our model is not consistent with this object being multiply imaged, and when including it as a constraint as multiply imaged (using spots in the arclike structure as independent images), the model predicts additional counter images which are not seen.  We therefore conclude that although this arc is lensed, it is not multiply imaged, and include it as a singly imaged system in the model constraints.

\section{Modeling}
\label{sec:modeling}
Following e.g., \citet{kneib1996, smith2005, limousin2007} we do a parametric mass reconstruction of Abell 2218.  The multiply imaged systems form the basis of our analysis, with each $n$-tuply imaged system giving $2(n-1)$ constraints if the redshift is known.  
As the number of constraints needs to be greater than the number of free parameters in our fit, we are limited by the known multiply imaged systems.  The MCMC sampling and optimization is done using the Lenstool software \citep{kneib1993, jullo2007}.   The optimization is performed in the source plane, as it is faster and has been found to be equivalent to optimizing in the image plane \citep{halkola2006, jullo2007}.  The new version of Lenstool \citep{jullo2007}, gives a distribution of values for each of the parameters, thus making it possible to estimate the uncertainty of the parameters.  In addition, it returns the Evidence of a model, which is a measure of how likely a model is, penalizing unnecessarily complicated models.  Thus, a model with a lower $\chi^2$ but more free parameters, may have a lower Evidence, suggesting that we should choose the simpler model.
For each image we find its rms (root-mean-square) value for its position in the source plane, rms$_s$, and image plane, rms$_i$, given by
\begin{eqnarray}
\mathrm{rms}_s&=&\sqrt{\frac{1}{n}\sum_{j=1}^{n}\left(X^j_s-<X^j_s>\right)^2}\\
\mathrm{rms}_i&=&\sqrt{\frac{1}{n}\sum_{j=1}^{n}\left(X^j_{\mathrm{obs}}-X^j\right)^2}
\end{eqnarray}
where $n$ is the number of images for the system, $X_s$ is the position in the source plane, $X$ the position in the image plane and $X_{\mathrm{obs}}$ the observed position in the image plane.  The overall rms is defined by summing and averaging over all the images for all the systems.
A detailed overview of the Lenstool software, including definitions of $\chi^2$ and the Evidence, can be found in \citet{jullo2007}.

\subsection{Model components}
\label{sec:dm_clumps}
\label{sec:scaling}
We refer to the individual components of the model as 'clumps', where each clump is denoted by its position, ellipticity, position angle and the parameters of the profile used to describe it.  The parametric profile we use is the dual Pseudo Isothermal Elliptical mass distribution (dPIE, derived from \citet{kassiola}) and its form and main properties are given in Appendix~\ref{app:piemd}.  The dPIE profile is defined in Lenstool by three parameters,  the core radius $a$, the scale radius $s$ and a fiducial velocity dispersion $\sigma_{\mathrm{dPIE}}$.  For $a<r<s$, the profile behaves as $\rho\sim r^{-2}$, while it falls like $r^{-4}$ in the outer regions.  For a vanishing core radius the scale radius corresponds to the radius containing half the 3D mass.

The clumps are called 'large scale clumps' if their mass within the outermost multiply imaged constraint is greater than $20\%$ of the total mass.  Smaller clumps are referred to as 'galaxy scale clumps', and are in general associated with the cluster members.  Large scale dark matter clumps are optimized independently.  The redshift is fixed at the location of the cluster, but the central position (R.A., Dec.), the ellipticity and the position angle (P.A.) are allowed to vary.  

We associate a galaxy scaled clump with each of the cluster galaxies, fixing the central location, ellipticity and position angle of the mass distribution to that of the light distribution.   A few of the cluster galaxies are fitted individually (optimizing their $a$, $s$ and $\sigma_{\mathrm{dPIE}}$, see Table \ref{tab:model} for an overview), but due to a lack of constraints, most of the galaxies are optimized in a combined way.  
The parameters ($a, s, \sigma_{\mathrm{dPIE}}$) are optimized together using the following scaling relations for the luminosity $L$:
\begin{eqnarray}
a=a^\star \left(\frac{L}{L^\star}\right)^{1/2}\\ s=s^\star \left(\frac{L}{L^\star}\right)^{1/2} \\  \sigma_{\mathrm{dPIE}}=\sigma_{\mathrm{dPIE}}^\star \left(\frac{L}{L^\star}\right)^{1/4}.
\label{eq:scaling}
\end{eqnarray} 
For a discussion of these scaling relations we refer to \citet{limousin2007} and \citet{halkola2007}.  For a given $L^\star$ luminosity, we fix $a^\star=0.25$~kpc, while $\sigma_{\mathrm{dPIE}}^\star$ and $s^\star$ are allowed to vary.   We note that fixing the core radius to be small, makes the profile approximately equivalent to the profile used by \citet{brainerd1996} to describe galaxies (see also Appendix~\ref{app:piemd}).  Following \citet{depropris1999} we take the apparent magnitude of an $L^\star$ in the K-band to be $K^\star=15$ at $z=0.171$ (redshift of Abell 2218).
\begin{deluxetable*}{lccccccc} 
\tablecolumns{8} 
\tablewidth{0pc} 
\tablecaption{Model parameters}
\tablehead{\colhead{Clump}    & \colhead{$\Delta$ R.A.} & \colhead{$\Delta$ Dec.} &   \colhead{$\hat{\epsilon}$}   & \colhead{$\theta_{\hat{\epsilon}}$} &\colhead{ $\sigma_{\mathrm{dPIE}}$} & \colhead{$a$} &\colhead{$s$}\\
&\colhead{$\arcsec$}&\colhead{$\arcsec$}&&& \colhead{(km/s)}& \colhead{($\arcsec$)}& \colhead{($\arcsec$)}}
\startdata 
DM1 &  $3.9^{+0.3}_{-0.4}$& $23.1^{+1.7}_{-0.7}$ & $0.033^{+0.03}_{-0.001}$&  $56^{+6}_{-6}$& $806^{+10}_{-10}$&$26^{+1}_{-1}$&$274^{+110}_{-4}$  \\
DM2 &   $-20.4^{+0.1}_{-0.6}$& $-23.1^{+0.6}_{-0.6}$ & $0.39^{+0.01}_{-0.01}$\tablenotemark{$\dagger$} &$ 8^{+1}_{-2}$  & $1029^{+13}_{-49}$&$44^{+2}_{-3}$&$462^{+85}_{-60}$\\
1193 (BCG) &  -0.5 & 0.07 & 0.46 &52.4  &$514^{+7}_{-11}$&$3.14^{+0.04}_{-0.45}$&$56^{+12}_{-8}$ \\
617 &   -46.1& -49.1 &  0.20& 59.4 &$288^{+7}_{-8}$&$2.5^{+0.3}_{-0.5}$& $87^{+10}_{-9}$\\
1028 &   -16.0&  -10.3& 0.18&  80.4&$396^{+13}_{-128}$&$0.3^{+0.2}_{-0.1}$& $0.6^{+0.5}_{-0.1}$\\
 &&&&& \colhead{$\sigma_{\mathrm{dPIE}}^\star$} & \colhead{$a^\star$} &  \colhead{$s^\star$} \\
&&&&& \colhead{(km/s)}& \colhead{($\arcsec$)}& \colhead{($\arcsec$)}\\
L$^\star$ galaxy & \nodata & \nodata & \nodata & \nodata & $185^{+10}_{-11}$& $0.09$ &$2.9^{+0.5}_{-0.3}$\\
\enddata 
\tablecomments{Values quoted without error bars were kept fixed in the optimization. The error bars correspond to $68\%$ confidence levels.  The location and the ellipticity of the matter clumps associated with the cluster galaxies were kept fixed according to the light distribution.  The center is defined at (R.A., Dec.)=(248.9546, 66.2122) in J2000 coordinates.}
\tablenotetext{$\dagger$}{The posterior probability distribution pushes towards the highest values of our input prior.  Therefore these error bars do not represent the full uncertainty in the value, but the uncertainty given a prior of $\hat{\epsilon}<0.4$  }
 \label{tab:model}
\end{deluxetable*}

\section{The Strong Lensing Mass Distribution} 
\label{sec:analysis}
\begin{deluxetable}{llll}
\tablecolumns{4} 
\tablewidth{0pc} 
\tablecaption{Goodness of the fit}
\tablehead{\colhead{System} & \colhead{$\chi^2$} &\colhead{rms$_s$ (\arcsec)} & \colhead{rms$_i$ (\arcsec)}}
\startdata
S1 &        21.7& 0.171 & 0.60\\
S2.1 &        10.0& 0.090 & 0.60\\
S2.2 &        11.8& 0.098 & 0.82\\
S2.3 &        37.9& 0.167 & 1.49\\
S2.4 &        55.8& 0.197 & 2.17\\
S3 &        4.9& 0.043 & 0.57\\
S4.1 &        9.8& 0.089 & 0.53\\
S4.2 &        2.1& 0.042 & 0.43\\
S5 &        15.6& 0.060 & 4.61\\
S6 &       0.3& 0.016 & 0.03\\
S7 &        8.4& 0.052 & 1.45\\
S8 &        0.0& 0.000 & 0.00\\
C1 &        13.4& 0.102 & 1.26\\
C2 &        20.5& 0.178 & 0.63\\
C3 &        17.5& 0.122 & 1.64\\
C4 &        4.7& 0.052 & 0.70\\
C5 &       0.2& 0.009 & 0.27\\
C6 &       0.5& 0.011 & 0.14\\
& \colhead{$\chi^2/\mathrm{d.o.f.}$} & \colhead{rms$_s$} & \colhead{rms$_i$}\\
Full model &        235.1/34 & 0.12 & 1.49\\
\enddata
\tablecomments{The $\chi^2$ and rms in the source plane (where the model is optimised) and the image plane.  The zero values found for S8 mean that the system is consistent with being singly imaged.}
\label{tab:chires}
 \end{deluxetable}

In this section we present our strong lensing model (optimized in the source plane), discuss its implications and compare it to previous results.  All reported error bars correspond to $68$\% confidence levels.  For our best model we find rms$_s=0\farcs12$, which gives rms$_i=1\farcs49$ (see Table~ref{tab:chires}).\footnote{A parameter file containing all the following information, and which can be used with the Lenstool software package, along with a FITS file of the mass map generated from the best-fit model are available at \url{http://archive.dark-cosmology.dk/}.  These can be used to find relevant critical lines for using Abell 2218 as a gravitational telescope and to predict/confirm candidate lensed systems. }

\subsection{A bimodal mass distribution}
\label{sec:mass_dist}
\begin{figure*}
\epsscale{.45}
\plotone{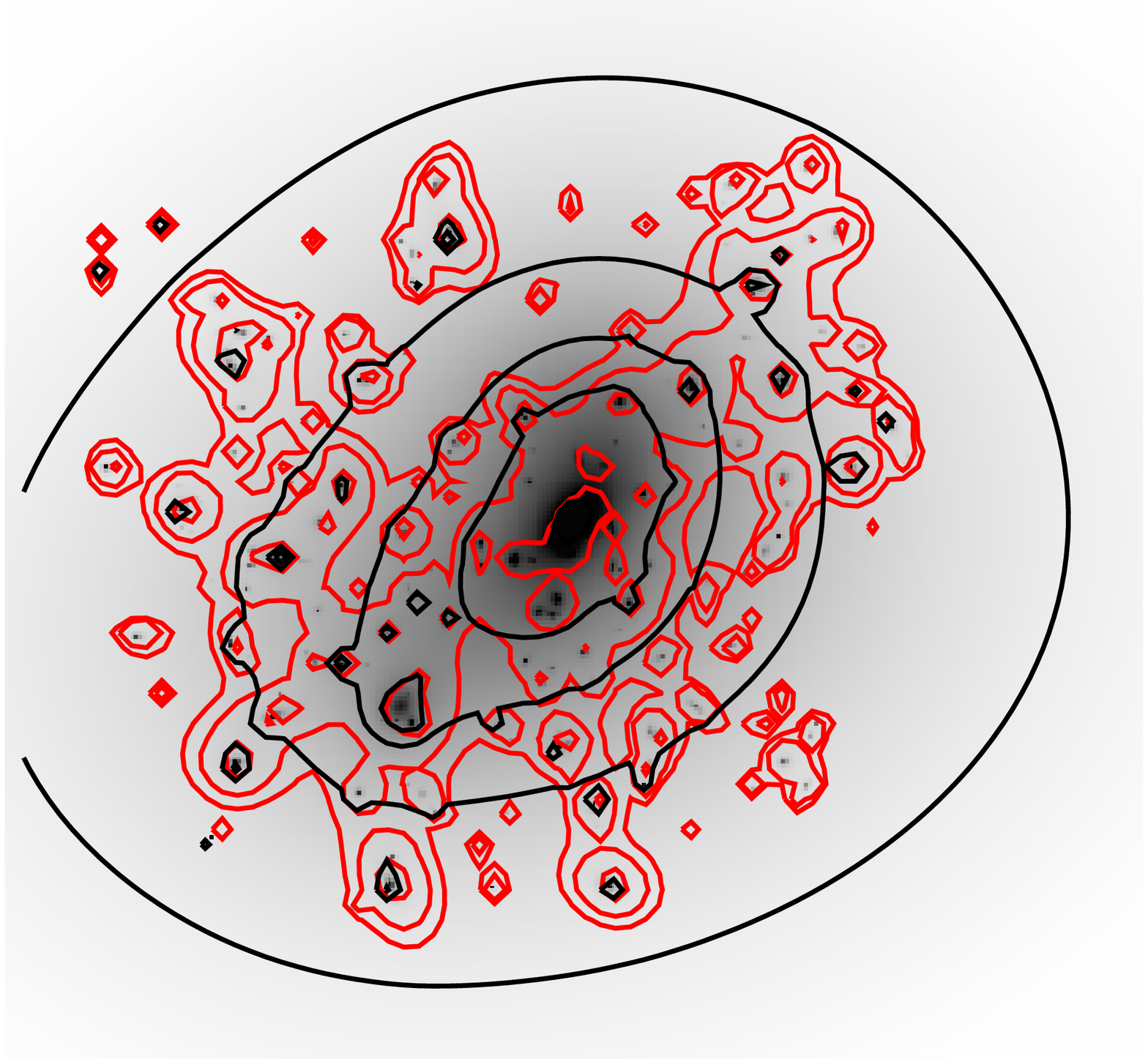}
\plotone{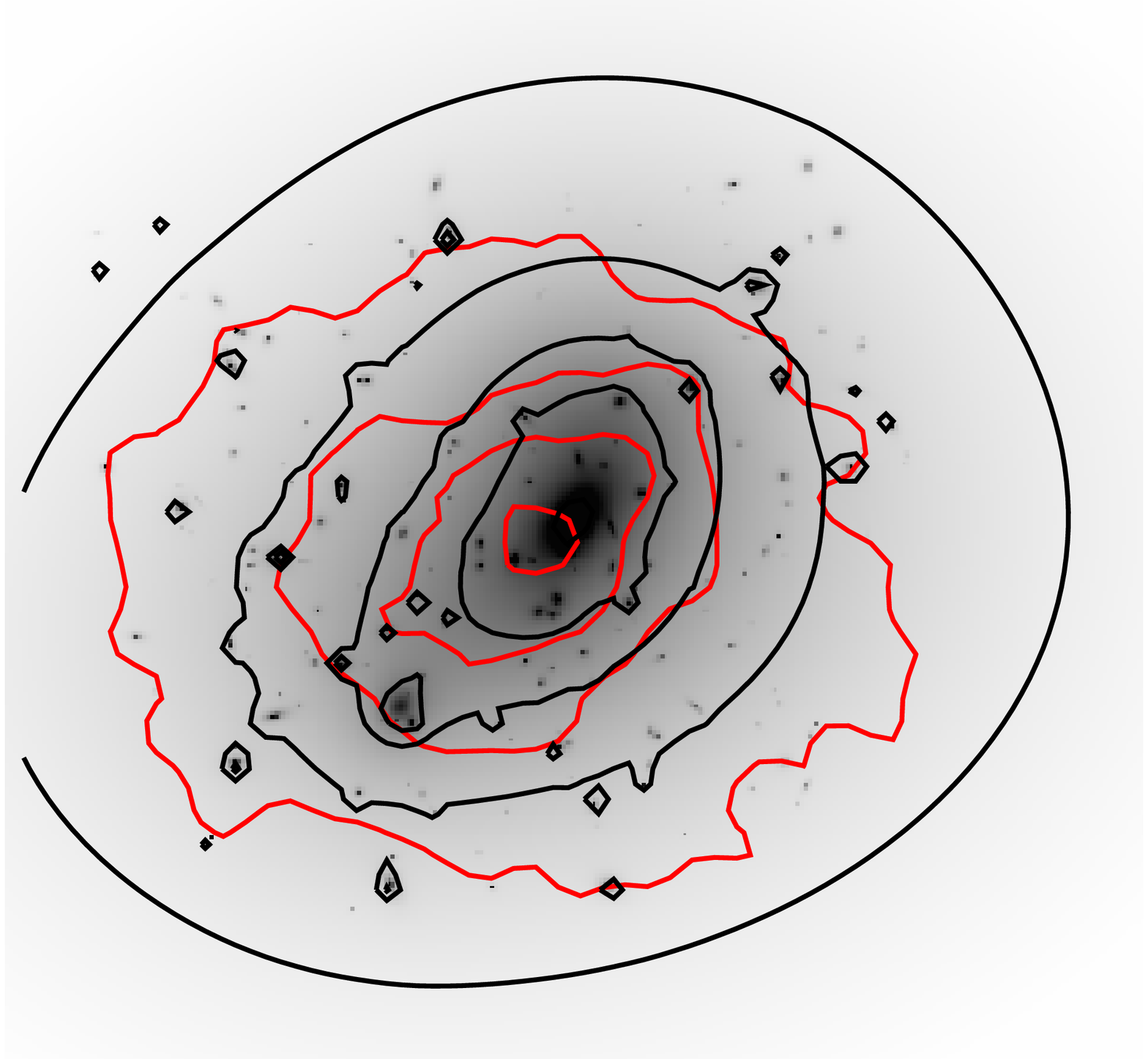}
\caption{The mass density map and its contours (black).  The maps are $300\arcsec\times300\arcsec$ and centered on the BCG with North being up and East being left.  The red contours show the light distribution (left panel) and the X-ray distribution (right panel).   The overall shape of the light-contours and the mass map contours are similar, with the light being slightly more 'pear'-shaped, as it broadens in the SE direction.  The overall shape of the light-contours and the mass map contours also agree, although the X-rays become more spherical for  in the outer regions.  See section~\ref{sec:bimodal}.  \label{fig:massmap}}
\end{figure*}
\begin{figure}
\epsscale{1.}
\plotone{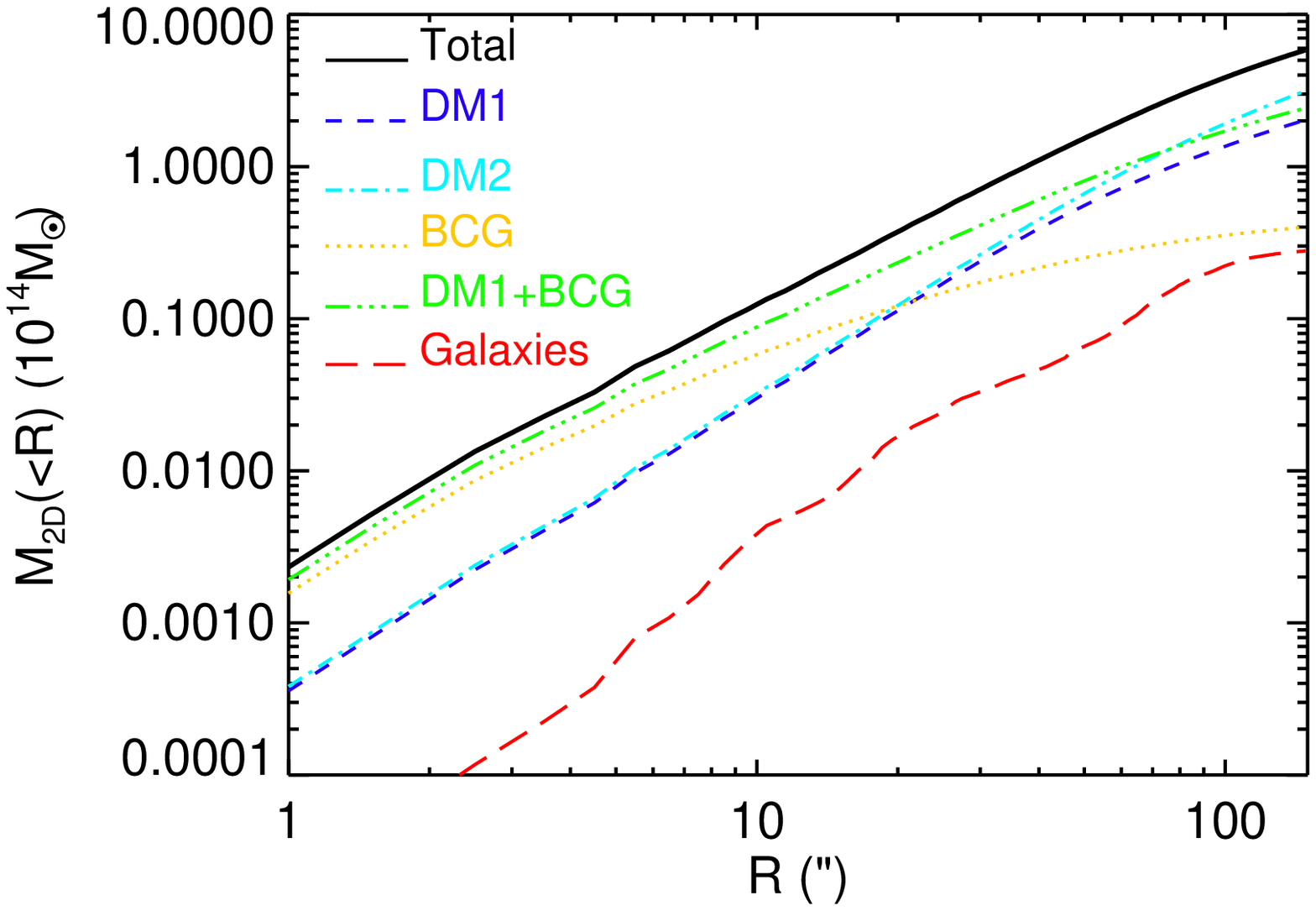}
\caption{Total projected mass as a function of aperture radius (centered on the BCG) for different model components.  The two large scale clumps, DM1 and DM2, contribute a similar amount to the mass, but the halo associated with the BCG dominates in the inner regions, and the combined halo of DM1 and the BCG dominates the mass in the region where most of the constraints lie (within $80\arcsec$).  The galaxies (excluding the BCG) only contribute a small amount to the overall mass, of the order of $5-6$\%. \label{fig:m_1d}}
\end{figure}

We show the mass density model in Figure~\ref{fig:massmap} and critical lines predicted by the model at $z=0.702, 2.515, 6.7$ in Figure~\ref{fig:overview}.  The total projected mass as a function of radius, centered on the BCG, is shown in Figure~\ref{fig:m_1d}.  
We find that the mass distribution is strongly preferred to be bimodal, even for a simple model where all the galaxies are modelled based on the scaling relations, but the dark matter clumps are optimized independently using only systems 1 and 2 as constraints (28 constraints in all).  The second dark matter clump is located to the south east of the central clump for this simple model, near galaxy \#617, which is in agreement with previous models of Abell 2218.  The bimodal mass distribution is also strongly preferred when we add more constraints.  We will refer to the two large scale dark matter clumps as DM1 (for the one associated with the BCG) and DM2 (for the one associated with galaxy \#617).  We also constructed a three clump model, but its bayesian Evidence was worse, leading us to reject it as our best model.

In previous models, the location of the second clump has been fixed at the center of the brightest galaxy in the south east corner (\#617), but we find that when the location is allowed to vary it is offset from this galaxy by $\sim 35\arcsec$.   
We also find that DM2 has high ellipticity and is comparable in mass to DM1, although significantly less massive than the DM1 and BCG halos combined (see Figure~\ref{fig:m_1d}).    We note that the light distribution is similar to the derived matter distribution (see Figure~\ref{fig:massmap}), supporting the finding that there is a significant matter component in the vicinity of DM2  (see also section~\ref{sec:bimodal}).

\subsection{Dark matter halos of galaxies}
\label{sec:dm_galaxies}
The potential contribution of the individual dark matter halos associated with cluster galaxies
was first proposed by \citet{natarajan1997}. Typically it is found that contribution of
dark matter halos associated with the bright, early-type cluster galaxies in the inner regions are required to explain the positions and geometries of multiply lensed images
in the strong lensing regime \citep{meneghetti2007}.

We fit three galaxies individually (for the parameters of these galaxies see Table~\ref{tab:model}), two are the brightest galaxies near the centers of the large scale dark matter clumps ($\#1193$ - the BCG - and $\# 617$), while the third is an elliptical galaxy important to the lensing of system S1.  The halo associated with the BCG is very massive, and in particular in the inner $10\arcsec$ it dominates the mass distribution (see Figure~\ref{fig:m_1d}).  Even at the outermost Einstein radii of a multiply imaged system ($80\arcsec$) it still contributes around $10\%$ of the mass of the cluster. 

System S1 has seven images, and two of the pairs are strongly affected by nearby galaxies (an elliptical $\# 1028$ and a spiral $\# 993$).  Keeping the parameters of the spiral free did not significantly affect the model, while adding the elliptical did.  Although both galaxies would at first glance appear equally important to be fitted individually due to their strong effects on system S1, \begin{figure}
\epsscale{1.}
\plotone{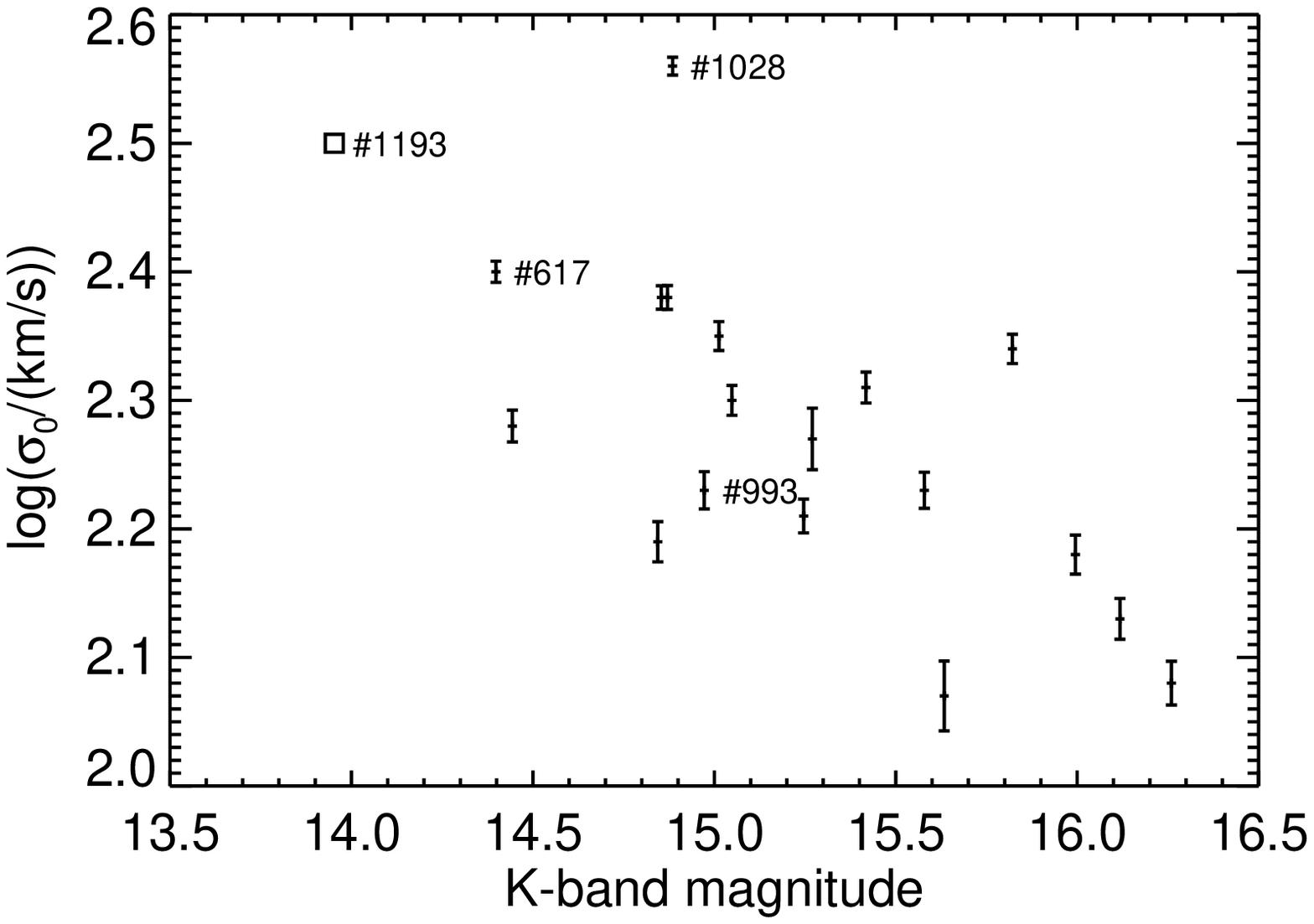}
\caption{The Faber-Jackson relation for the K-band using $\sigma_0$ measurements from \citet{ziegler2001}.  Also plotted is the central velocity dispersion of the BCG ($\# 1193$) as obtained by \citet{jorgensen1999}.   The individually marked galaxies are discussed in section~\ref{sec:scaling} and \ref{sec:dm_galaxies}. \label{fig:fab_jack}}
\end{figure}
Figure~\ref{fig:fab_jack} provides an important insight to why only $\# 1028$ needs to be included individually.  This is because the scaling relations (see \S~\ref{sec:scaling}) assume that the galaxies follow a Faber-Jackson relation (Equation~\ref{eq:scaling}) without any scatter.  As can be seen from Figure~\ref{fig:fab_jack} and  discussed in \citet{ziegler2001} the galaxies in Abell 2218 show a significant scatter and therefore the above scaling relations can give very unrealistic values for individual galaxies which lie far from the mean behavior.  The elliptical $\# 1028$ is one such galaxy, with a measured central velocity dispersion which is greater than the central velocity dispersion of the BCG as measured by \citet{jorgensen1999}, while $\# 993$, although a spiral galaxy, is more consistent with the mean.

For the other cluster galaxies, they are included in the model via the scaling relations given in Section~\ref{sec:scaling}.  For $L^{\star}$ corresponding to $K=15$ at $z=0.171$ we find $\sigma_{\mathrm{dPIE}}^\star=185^{+10}_{-11}$~km~s$^{-1}$ and $s^\star=2\farcs9^{+0.5}_{-0.3}$, but we note that there is some degeneracy between the two values (see discussion in section~\ref{sec:degeneracy}).

\subsection{Comparison with measured velocity dispersions}
We use the velocity dispersion measurements, $\sigma_0$, of \citet{ziegler2001}  and \citet{jorgensen1999} for Abell 2218 to compare with the results from the lens model.   \citet{ziegler2001} have a total of 48 galaxies in their sample, of which nearly half fall within the ACS image and are included in our galaxy catalogue, while the  \citet{jorgensen1999}  data contains the BCG and seven other cluster members in our galaxy catalogue.  Using the velocity dispersion measurements for Abell 2218 cluster members from \citet{ziegler2001} we plot the Faber-Jackson relationship for the K-band data in Figure~\ref{fig:fab_jack}.  We then calculate the mean and the standard deviation of galaxies with K-band magnitudes in the range from $14.8$ to $15.2$, excluding the individually fitted galaxy (galaxy $\# 1028$ in our catalog, or $\# 1662$ in the notation used by \citet{ziegler2001}).  This gives $\sigma_{0,\mathrm ziegler}^\star\approx195\pm35$~km~s$^{-1}$ for a typical K=15 galaxy.

\begin{figure}
\epsscale{1.}
\plotone{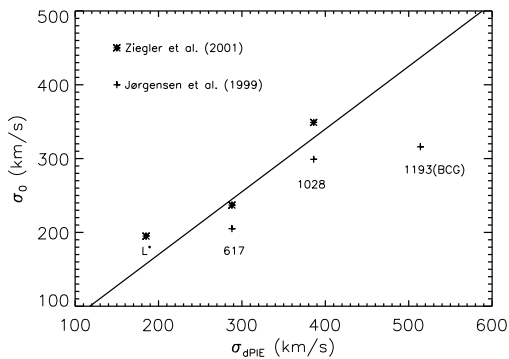}
\caption{The velocity dispersion of the dPIE, $\sigma_{\mathrm{dPIE}}$ vs. $\sigma_0$ measurements from \citet{ziegler2001} and \citet{jorgensen1999}.  The line is not a fit, but shows the relationship $\sigma_0=0.85\sigma_{\mathrm{dPIE}}$ found in Section~\ref{sec:app_veldisp}. \label{fig:compare_vel_disp}}
\end{figure}
In Figure~\ref{fig:compare_vel_disp} we plot $\sigma_0$ from \citet{ziegler2001} and \citet{jorgensen1999} vs. the fiducial velocity dispersion, $\sigma_{\mathrm{dPIE}}$ for the BCG, the two individually fitted galaxies and $L^{*}$.  
 Although the values from direct velocity dispersion measurements can not be directly related to the values obtained from the dPIE profile (as the measured values are calculated based on an isothermal profile and are not aperture corrected) we find that they are consistent with $\sigma_{\mathrm{dPIE}}$ being related to the measured velocity dispersion by $\sigma_0\approx0.85\sigma_{\mathrm{dPIE}}$ as found in Section~\ref{sec:app_veldisp}.

\subsection{Redshift estimates of the new candidate systems}
\label{sec:redshifts}
We estimated the redshifts of the new candidate systems using the model predictions.  The three component candidate system C2 is found to have $z=2.6\pm0.1$.  For the merging arcs we find $z=2.8\pm0.6$ for C3, $z=2.2\pm0.2$ for C4,  $z=2.6\pm0.3$ for C6 while C5 is poorly constrained with $z=2.3\pm0.8$. The estimated redshifts are reported in Table~\ref{tab:systems}.  These redshifts are consistent with a preliminary photometric redshift analysis done using the Hyperz photometric redshift code \citep{bolzonella2000}.

\subsection{A strongly lensed galaxy group at z=2.5}
\begin{figure}
\epsscale{1.0}
\plotone{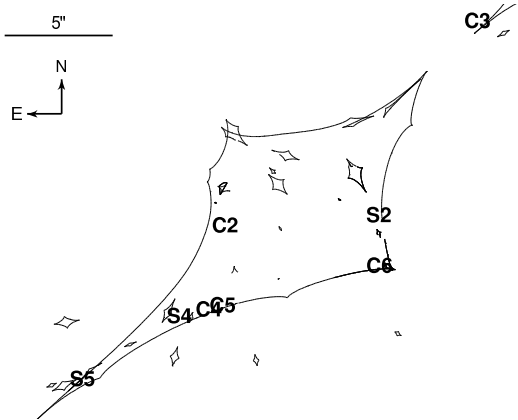}
\caption{A plot of the source plane at $z=2.515$ showing systems S2, S4 and S5 and the caustic lines.  Their maximum separation is around $130$~kpc, consistent with them belonging to the same group of galaxies.  Also shown are the candidate systems C2, C3, C4, C5 and C6 when their redshift is assumed to be $z=2.515$.  Under that assumption, their locations are consistent with them also belonging to this group of galaxies.\label{fig:2515}  }
\end{figure}

Three multiply imaged systems, S2, S4 and S5, all have the same redshift of $z=2.515$.  To check whether these three systems belong to a background galaxy group, we lens them back to their source plane (see Figure~\ref{fig:2515}).  

We find that S4 is in the middle with S5 and S2 at a separation of $5\farcs4$ and $10\farcs3$ respectively, with the maximum separation being $15\farcs7$, corresponding to $127$~kpc in the source plane ($1\arcsec = 8.06$ kpc at $z=2.515$), suggesting that the three systems belong to the same background group of galaxies.  This is in agreement with the findings of \citet{kneib2004b} who found the maximum separation of the systems in the source plane to be $130$~kpc.  It may be of interest to do a dedicated search for more systems at $z=2.515$, either multiply imaged or singly imaged, to further study this high redshift group of galaxies, and we note that all the candidate systems have a predicted $z$ consistent with $z=2.515$.  If we assume that these candidate systems have $z=2.515$, their location in the source plane is consistent with them belonging to this same group (see Figure~\ref{fig:2515}).

\subsection{Comparison with previous results and weak lensing}
Although the overall results of our model are in agreement with previous models of Abell 2218, they have found the second clump to be less massive \citep[see e.g.,][]{kneib1996, abdelsalam1998, smith2005,natarajan2007}.  There are several possible reasons for the discrepancy, with the first one being the new constraints we use in this model.  We have therefore redone the modeling using only the previously known spectroscopically confirmed systems, but we still find that DM2 is massive and with a large core.  It is also possible, that the previous models constructed using the older version of the lenstool package which did not involve MCMC sampling, have been caught in local minima.  Indeed, we do find when forcing DM2 to be smaller, a comparable $\chi^2$ but with the posterior probability distribution of the core radius $a$ pushing toward the upper limit of the input range.  However, this can not explain the discrepancy found for the non-parametric model of \citet{abdelsalam1998}.  

\begin{figure}
\epsscale{1.0}
\plotone{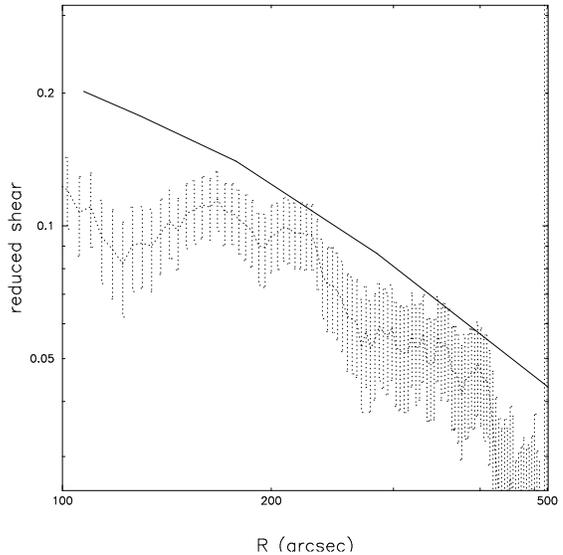}
\caption{ The weak lensing signal predicted by our model (solid line) compared to the weak lensing found by \citet{bardeau2007} (dotted line - $1\sigma$ error bars).   The \citet{bardeau2007} shows a very flat inner profile, characteristic of contamination of the background galaxy catalog by cluster members.  Therefore, we do not expect an agreement in the inner regions.  In the outer regions, where the contamination should be negligible, we find that our model overpredicts the weak lensing signal, but we note that the prediction is an extrapolation of a strong lensing model with constraints within $100\arcsec$.  \label{fig:weak_lens}  }
\end{figure}

A third possible explanation is that the models of  \citet{abdelsalam1998, smith2005,natarajan2007} all incorporated weak lensing to 'anchor' the outer part of the mass distribution.  Using our model to predict the weak lensing shear profile at large radii, we find that our profile overestimates the signal compared to the measured signal of \citet{bardeau2007} from ground based observations with the Canada-France-Hawaii Telescope (see Figure~\ref{fig:weak_lens}).  
The central part of the \citet{bardeau2007} profile is flat, characteristic of contamination of the background galaxy catalog by foreground cluster members (see \citet{limousin2007} for discussion on contamination).  Therefore, we do not expect to have an agreement between the weak lensing result and the strong lensing result.  At around $\sim300-400\arcsec$ the contamination should be negligible, and there we find that the agreement is better although our model still overpredicts the signal (although they are consistent within $2\sigma$).  We stress however that at this radius, we are extrapolating a strong lensing model based on constraints within $100\arcsec$, and the prediction becomes more uncertain the further we go out.

\section{Degeneracies}
\label{sec:degeneracy}
In Section~\ref{sec:analysis} we presented the strong lensing model of Abell 2218.
In this Section we study the degeneracies of our parametric strong lensing modeling, both those inherent to the parametric profile and the model components.  \citet{jullo2007} have also discussed the various degeneracies of the dPIE profile in lensing, addressing how different image configurations can break some degeneracies.

Lensing most strongly constrains the projected mass, and therefore we expect to see degeneracies arising from Equation~\ref{eq:mass2D}, although ellipticity may complicate that picture further.   In agreement with \citet{jullo2007} we find for the large scale halos, that the scale radius, $s$, is poorly constrained (as it lies beyond the outermost multiply imaged system).  This is also the case for the BCG and the \#617 which have large scale radii.  For the smaller halos, i.e., the scaled galaxies and \#1029, the scale radius, s, is small enough to affect the projected mass, and we find that lower $s$ requires higher $\sigma_{\mathrm{dPIE}}$ to keep the mass constant 
\begin{figure}
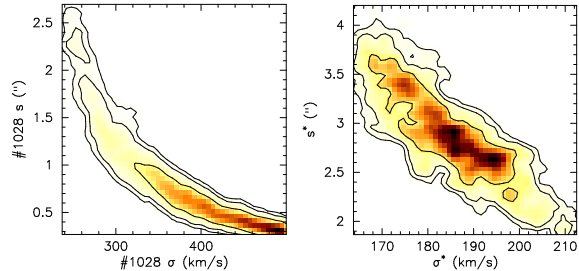

\epsscale{.50}
\plotone{f13a.eps}
\plotone{f13b.eps}
\caption{The degeneracy between the scale radius, $s$, and $\sigma_{\mathrm{dPIE}}$ for \#1028 (left) and the scaled galaxies (right).  These arise from equation \ref{eq:mass2D} which gives the aperture mass, showing that to keep the enclosed mass constant, an increase in the scale radius requires a lower value of $\sigma_{\mathrm{dPIE}}$.  The contours correspond to $1\sigma$, $2\sigma$ and $3\sigma$ confidence levels.  \label{fig:sigma_scale}  }
\end{figure}
(see Figure~\ref{fig:sigma_scale}).  However, the favored region for $s$ is always small, consistent with the tidal stripping of cluster galaxies proposed by \citet{natarajan1998,natarajan2002a,natarajan2002b,limousin2007AA, limousin2007_trunc}.

\begin{figure*}
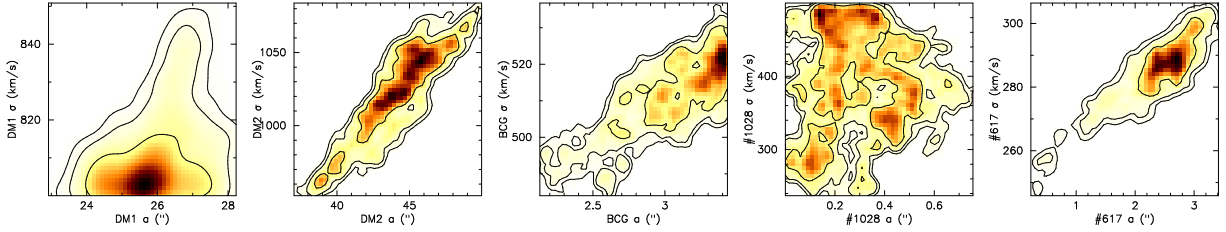

\epsscale{.20}
\plotone{f14a.eps}
\plotone{f14b.eps}
\plotone{f14c.eps}
\plotone{f14d.eps}
\plotone{f14e.eps}
\caption{The degeneracy between the core radius, $a$, and $\sigma_{\mathrm{dPIE}}$ for DM1, DM2, BCG, \#1028, \#617 from left to right.  These again arise from Equation~\ref{eq:mass2D} requiring the aperture mass to remain constant.  They show that to keep the mass constant, the $\sigma_{\mathrm{dPIE}}$ must increase when the core radius $a$ is increased.  This is consistent with the findings of \citet{jullo2007} for the dPIE profile and the findings of \citet{kochanek1996} for more general profiles with core.  The contours correspond to $1\sigma$, $2\sigma$ and $3\sigma$ confidence levels.\label{fig:sigma_core}}
\end{figure*}
Also, in agreement with \citet{jullo2007}, and as discussed by e.g. \citet{kochanek1996} for more general cored profiles, a larger core radius, $a$,  requires higher $\sigma_{\mathrm{dPIE}}$ to keep the mass constant (see Figure~\ref{fig:sigma_core}).  As our model has both a large core radius $a$ and $\sigma_{\mathrm{dPIE}}$ for DM2 compared to previous models of A2218, we have explored whether this degeneracy can reduce both values.  However, when forcing both $a$ and $\sigma_{\mathrm{dPIE}}$ to be smaller, the posterior distribution of $a$ always pushes to the upper limit of the input range.  Therefore, we conclude that this degeneracy is not the explanation for the high $a$ and $\sigma_{\mathrm{dPIE}}$ we find for DM2.  This flat profile is further supported by the 'blind tests' we perform in Section~\ref{sec:blind_test}.

In addition to the degeneracies associated with the profile itself, there may be degeneracies associated with the components we include in the model, i.e., are all the components we include necessary and are they independent of each other?  As mentioned in section~\ref{sec:mass_dist}, the model is strongly preferred to be bimodal.   
\begin{figure}
\epsscale{.5}
\plotone{f15a.eps}
\plotone{f15b.eps}\\
\plotone{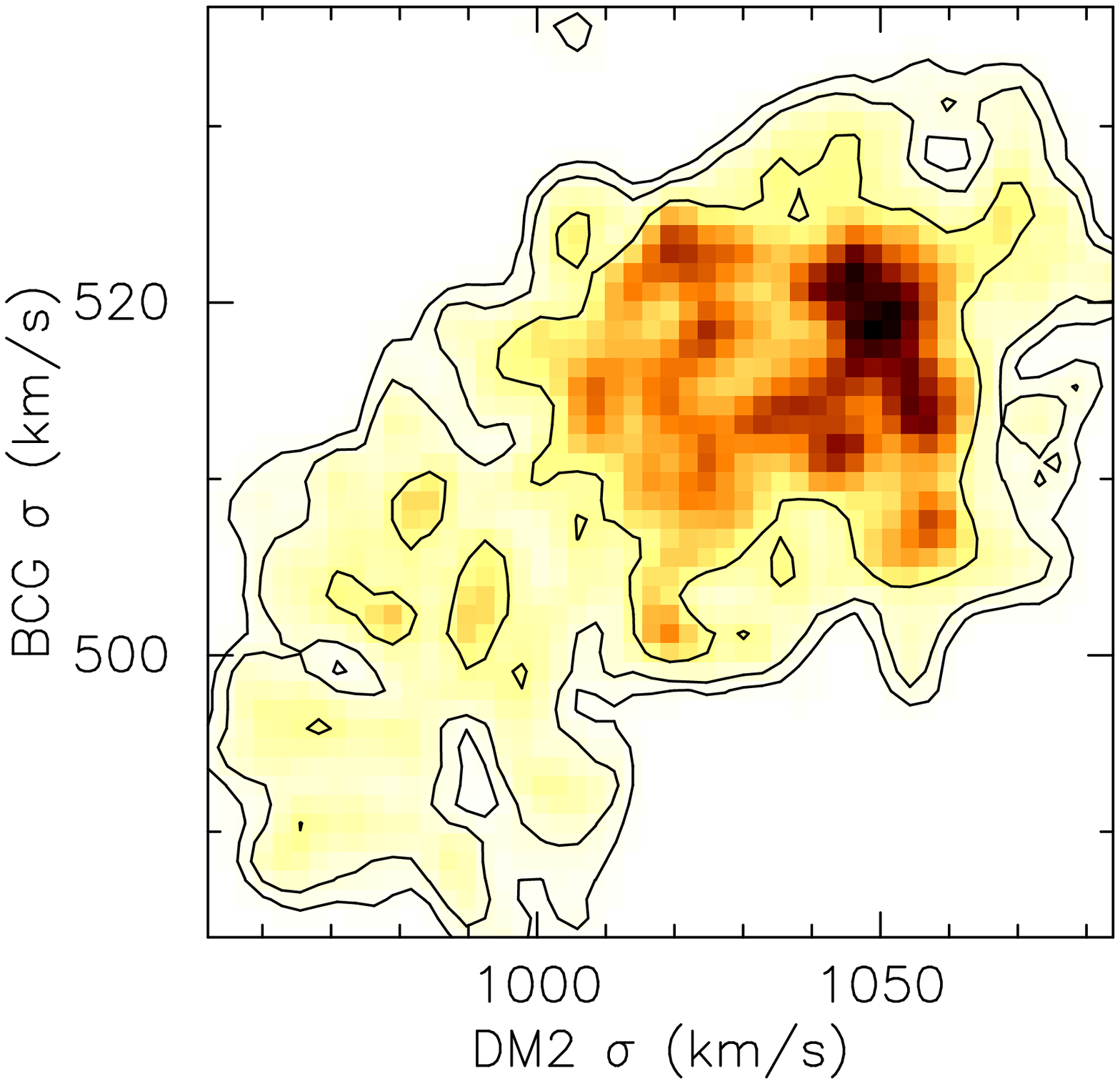}
\plotone{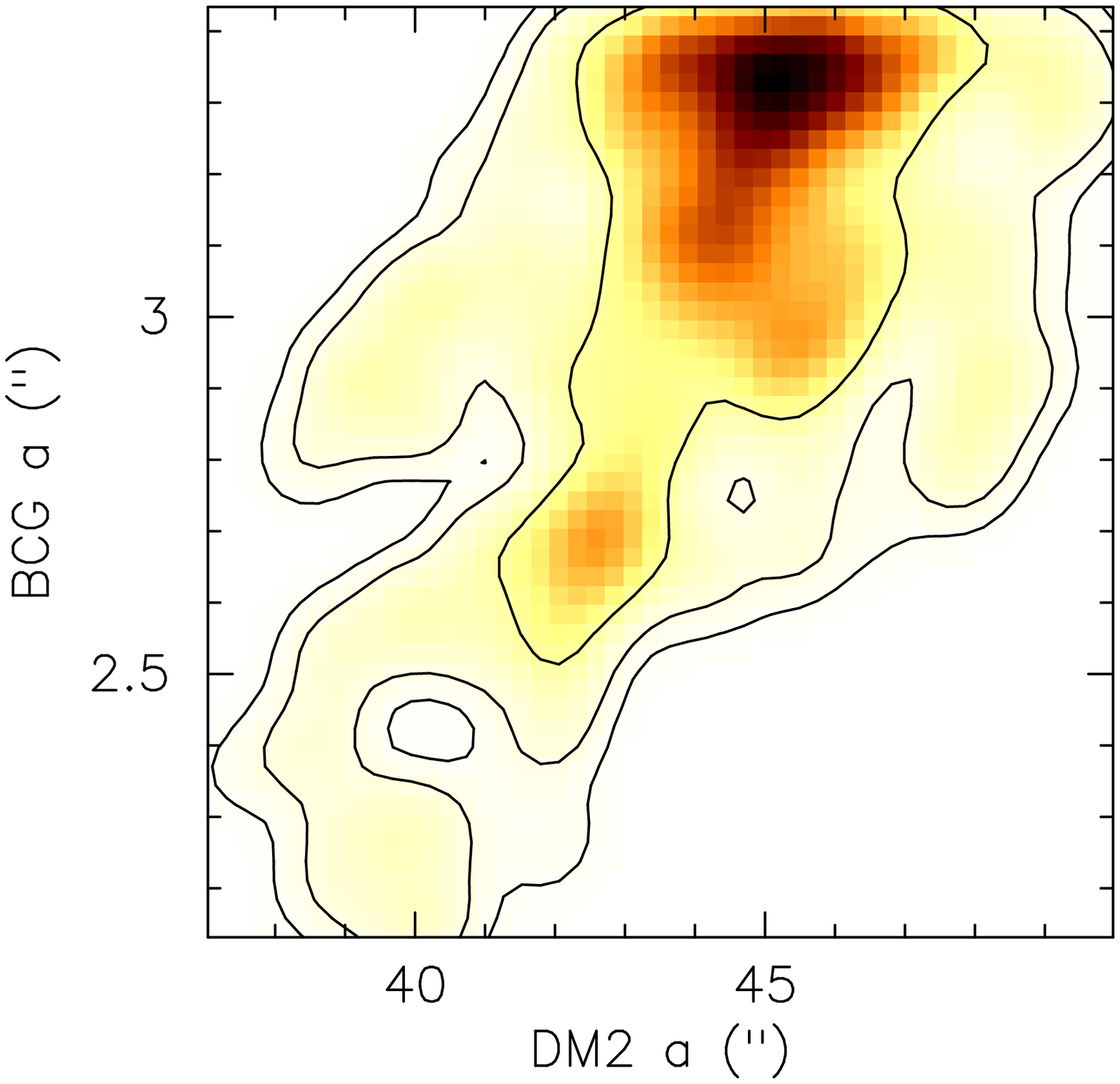}\\
\plotone{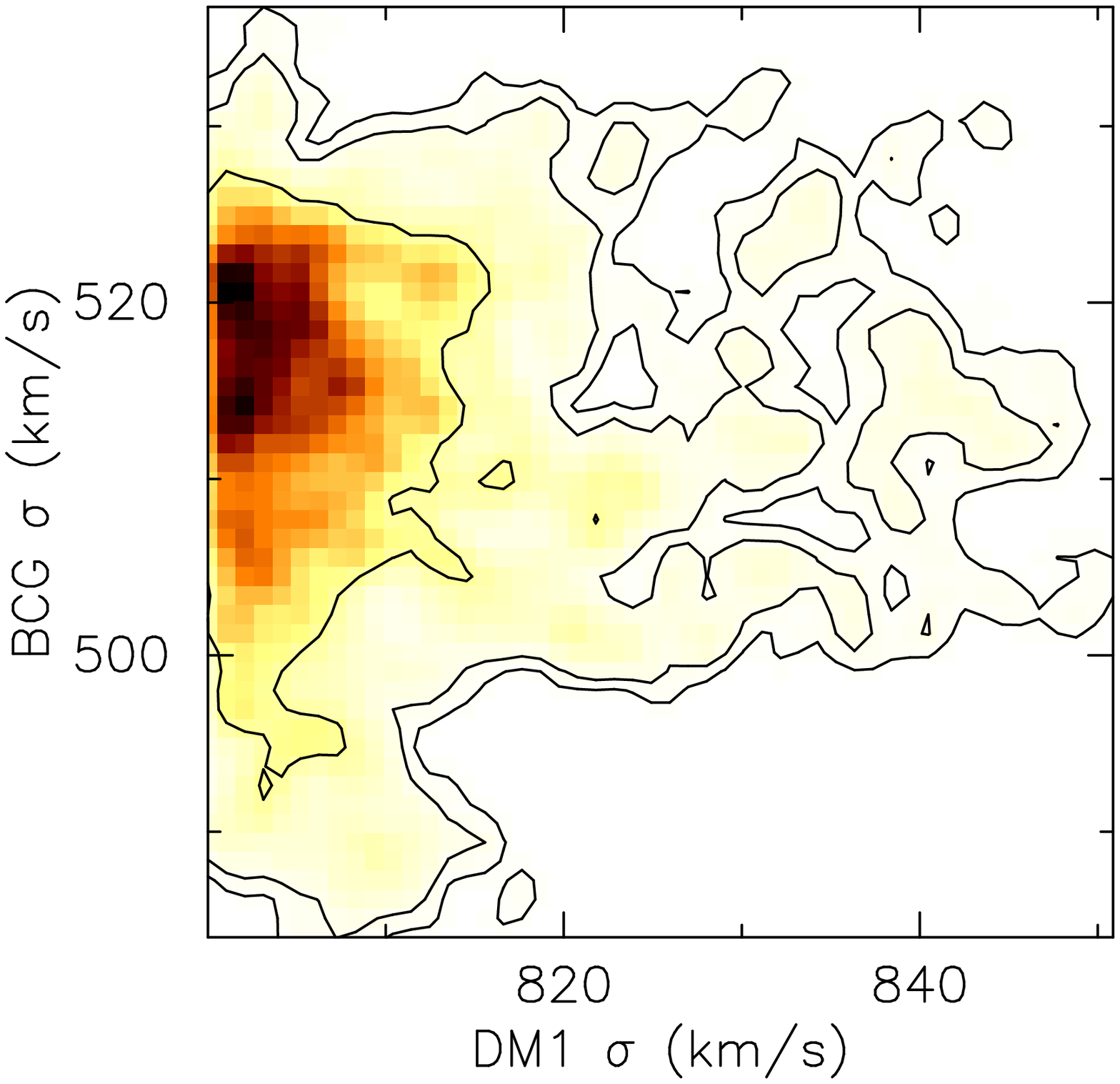}
\plotone{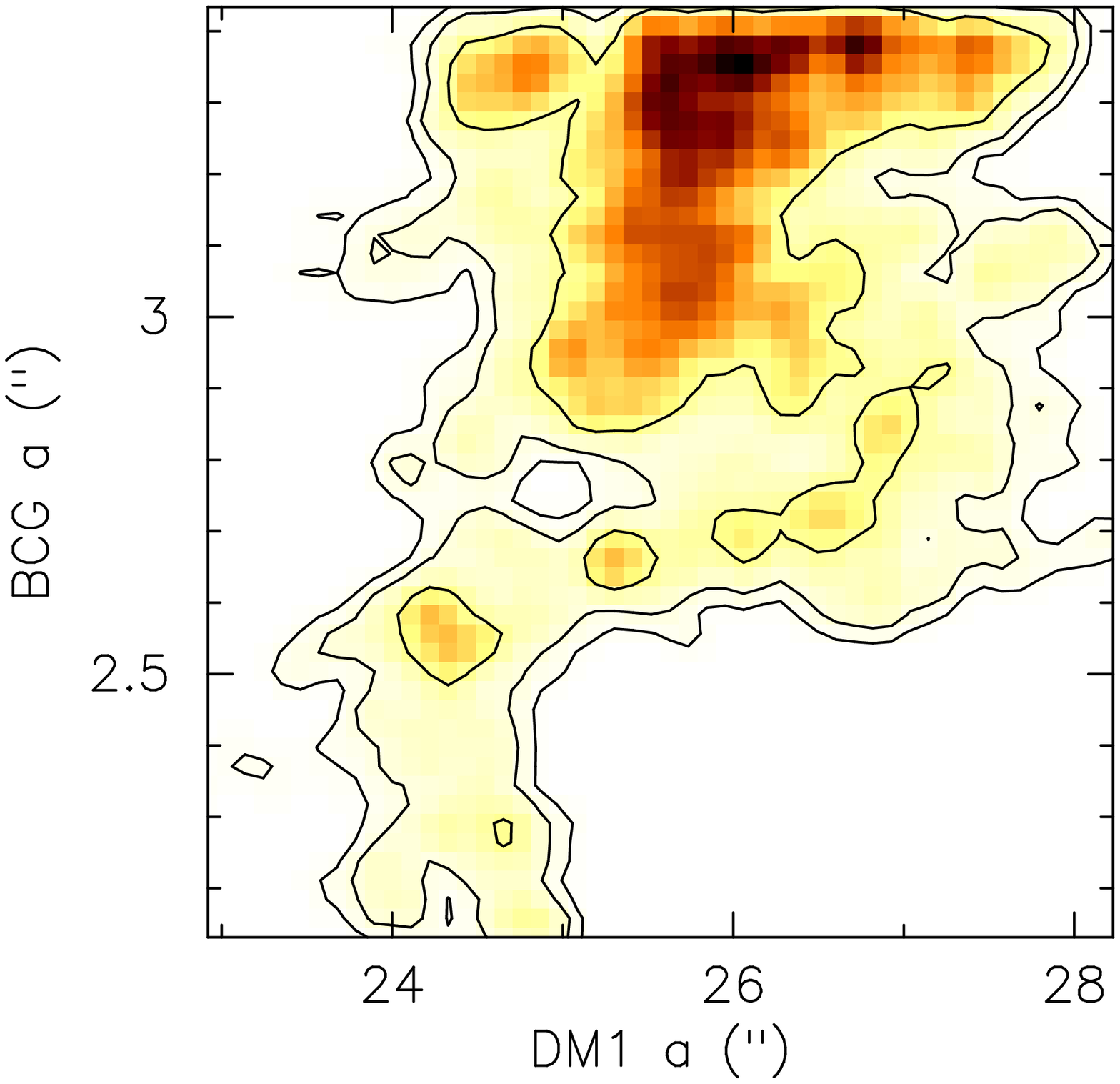}
\caption{The 2D posterior distribution of the parameters of DM2 vs. DM1 and BCG.  There is a clear degeneracy between the parameters of the two large scale components and the BCG.   Therefore, although the model prefers the inclusion of all the components, the values of their parameters are not fully independent.  The contours correspond to $1\sigma$, $2\sigma$ and $3\sigma$ confidence levels.\label{fig:DM1vsDM2deg}  }
\end{figure}
There is however degeneracy between the parameters of DM1 (and the BCG) and DM2 which is visualized in Figure~\ref{fig:DM1vsDM2deg}.  As for the BCG, we find that the Evidence is marginally higher when both the DM1 and BCG  are included in the model, suggesting that the data are sufficient to model both separately. Therefore we conclude, that although the model prefers the inclusion of all three components, the values for their parameters are not fully independent.  Finally, we look at the parameters for the scaled galaxies with respect to DM1 and DM2 
\begin{figure}
\epsscale{.5}
\plotone{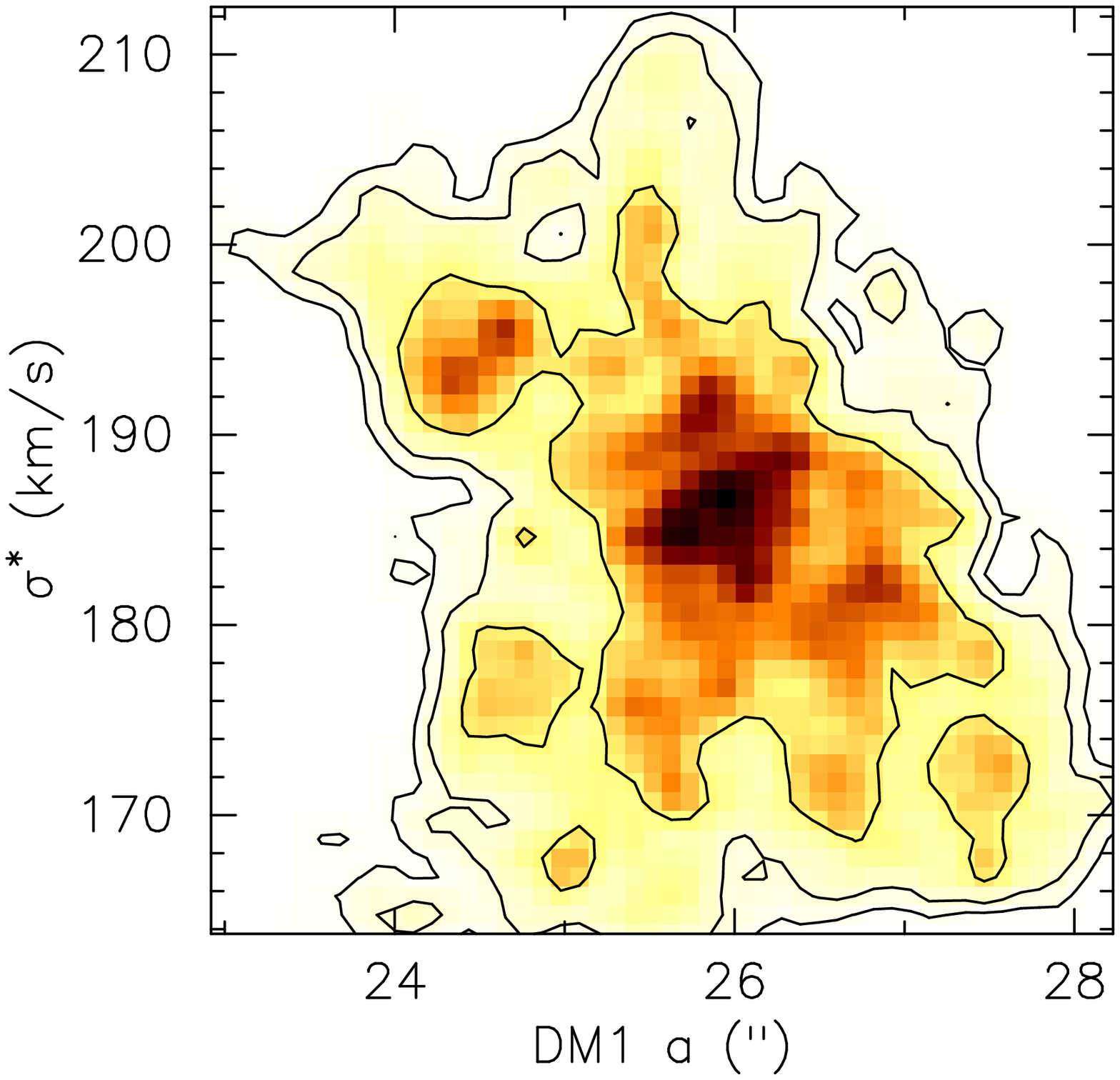}
\plotone{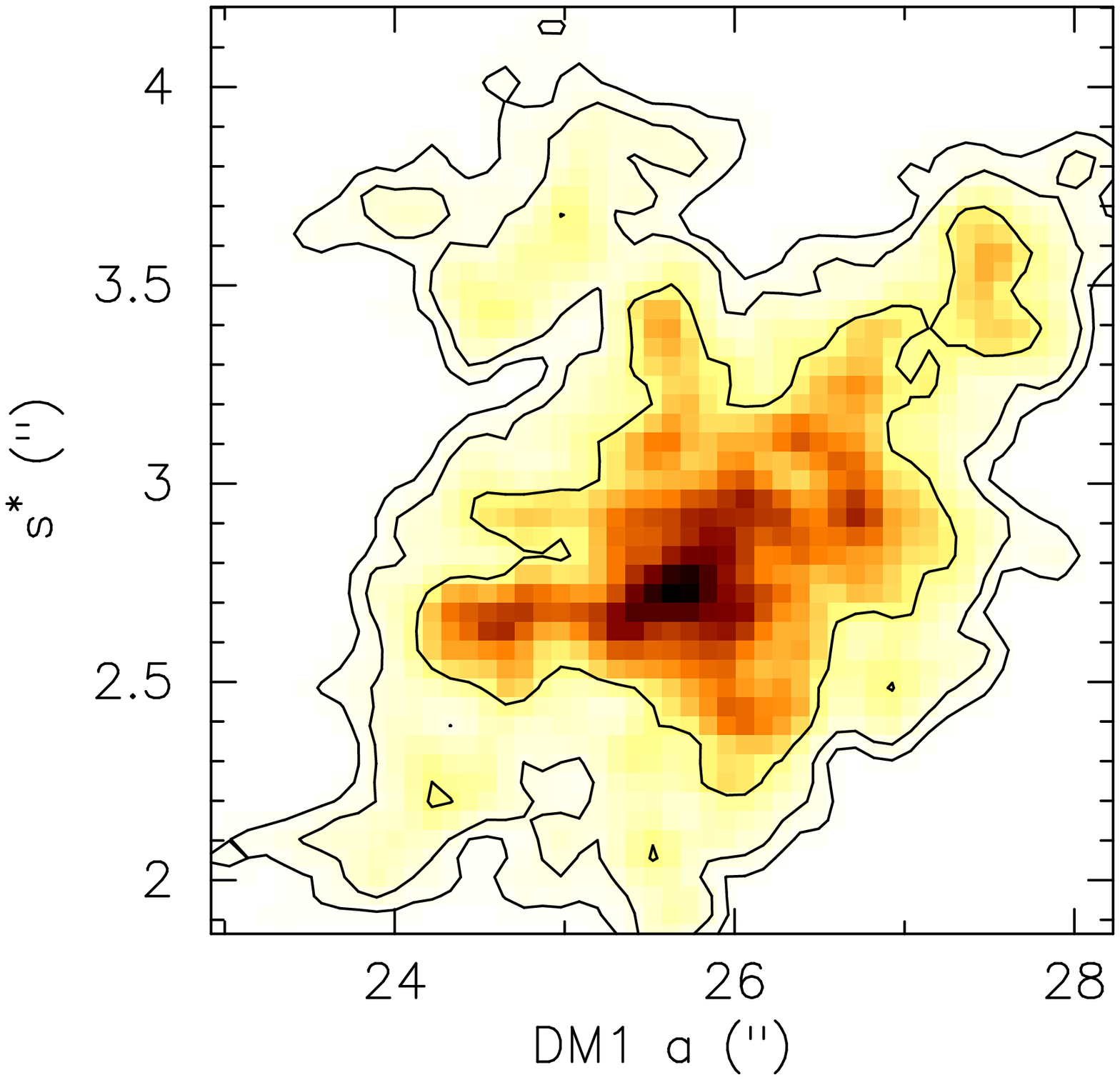}\\
\plotone{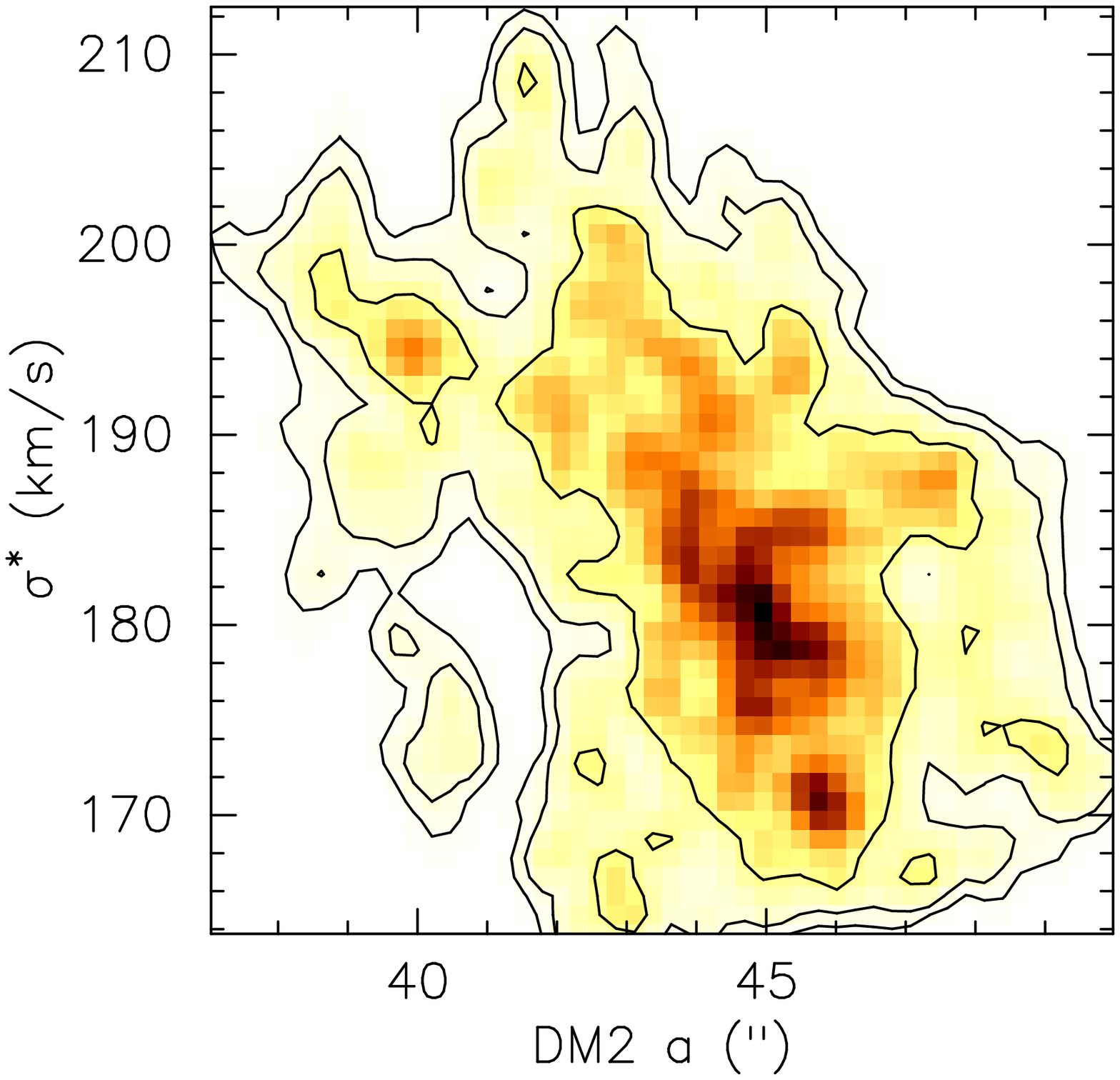}
\plotone{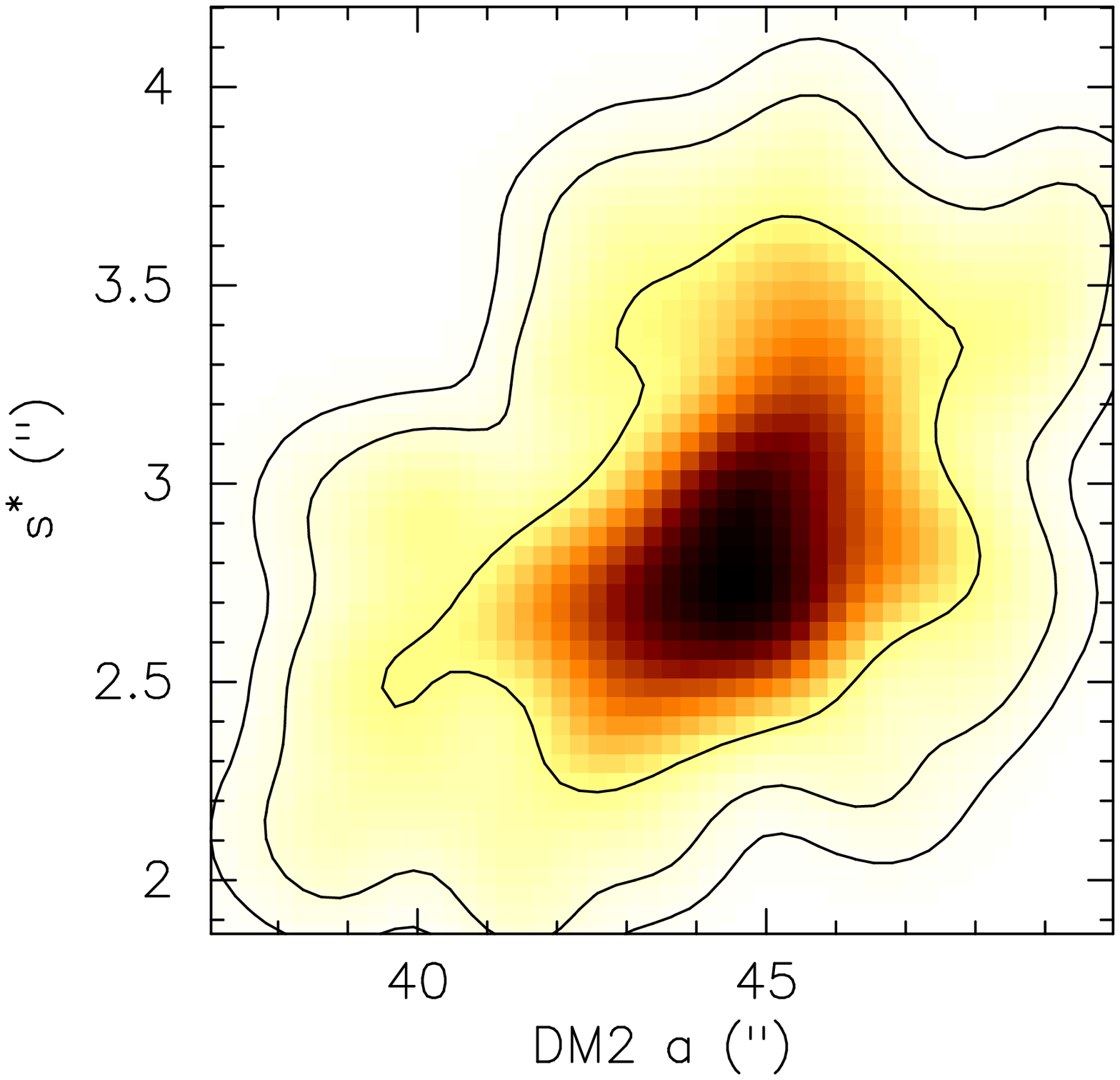}
\caption{The 2D posterior distribution of the scaled galaxy parameters ($\sigma^\star$, $s^\star$) with respect to the core radii of the large scale dark matter clumps DM1 and DM2.   Although there is a degeneracy between the parameters, the scale radius $s^\star$ is well constrained at low values.  The contours correspond to $1\sigma$, $2\sigma$ and $3\sigma$ confidence levels.\label{fig:POTvsDMdeg}  }
\end{figure}
(see Figure~\ref{fig:POTvsDMdeg}).  Although there is degeneracy the preferred value, in particular for the scale radius $s^\star$, is well constrained at low values ($\lesssim4\arcsec$), consistent with the tidal stripping scenario.

\begin{figure*}
\epsscale{.5}
\plotone{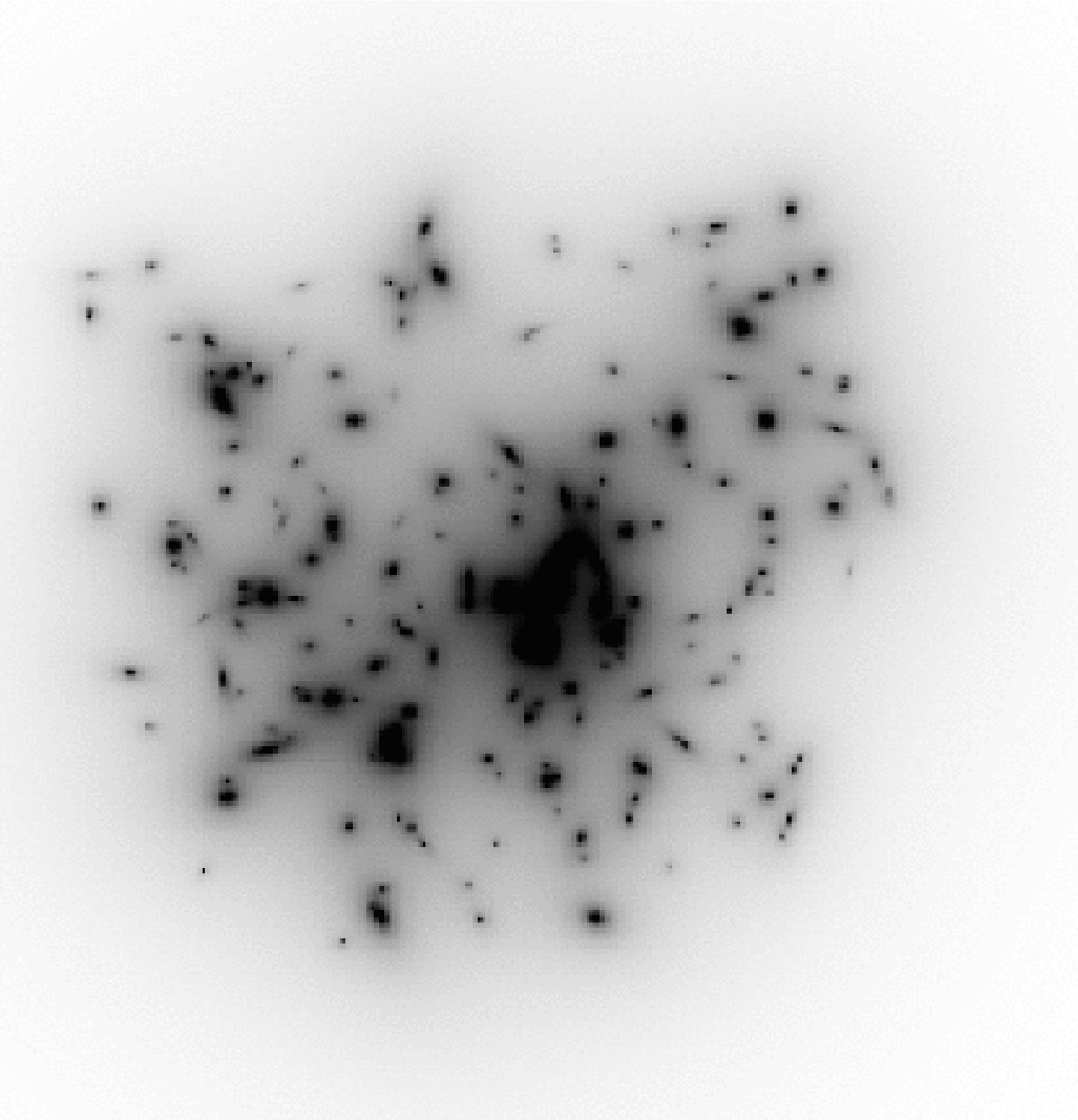}
\plotone{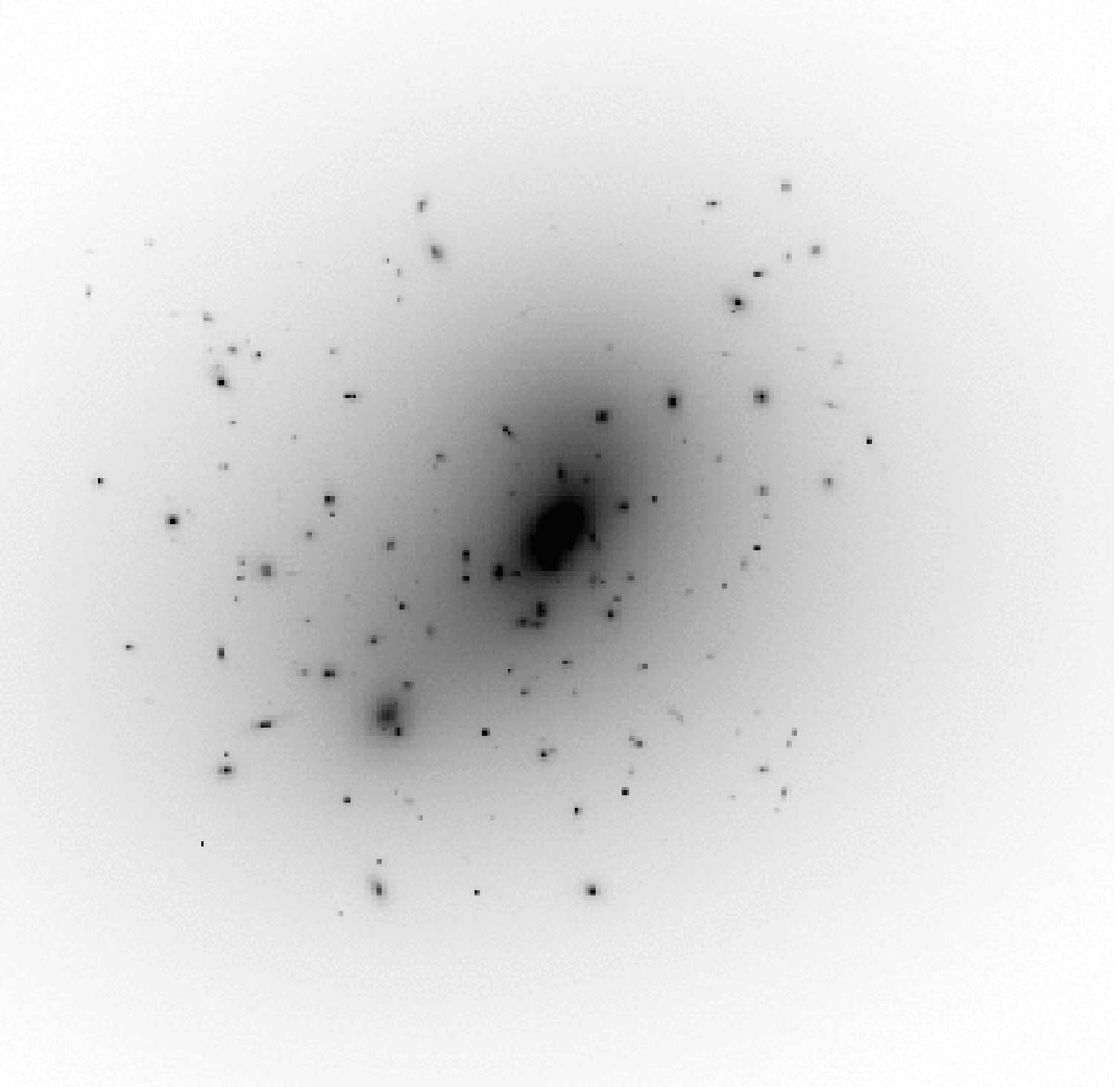}
\caption{ {\it Left panel:} The mass map from Section~\ref{sec:need_DM} using only halos associated with the galaxies.   {\it Right panel:} The mass map from Section~\ref{sec:analysis} consisting of large scale dark matter halos and galaxy halos.  The mass map including the large scale dark matter halos, which is much smoother, provides a significantly better fit to the data than the clumpy galaxy-only model (rms$_s=0\farcs12$ vs. rms$_s=1\farcs62$).  The maps are $300\arcsec\times300\arcsec$ and are centered on the BCG with North being up and East being left. \label{fig:need_DM}  }
\end{figure*}
\section{Reliability of the mass map}
\label{sec:reliability}
We have presented the mass map inferred from our strong lensing analysis in Section~\ref{sec:analysis} (Figure~\ref{fig:massmap}).
In short, we find evidence for a bimodal mass distribution described by two large scale smooth dark matter clumps, on top of which
we add some extra mass associated with the cluster galaxies.
Moreover, we associate a significant mass concentration with the location of the BCG galaxy.
These two conclusions are believed to be robust, and the aim of this section is to perform a series of tests in order to check 
our main findings regarding how the dark matter is distributed in Abell~2218.

\subsection{Do we need the smoothly distributed dark matter component?} 
\label{sec:need_DM}
We wish to test whether the smoothly distributed component is necessary to reproduce a good fit, or whether mass associated with the galaxies alone can provide an equally good fit.  To this end,
we construct a model where the mass is only in halos associated with the galaxies, without the inclusion of any smoothly distributed dark matter component.  The galaxies are included in a scaled manner, allowing the $\sigma_{\mathrm{dPIE}}^\star$ and $s^\star$ to move to higher values to increase their mass.
As a result of the optimization, we find a very poor fit (rms$_s=1\farcs62$ instead of $0\farcs12$ for the model from Section~\ref{sec:analysis}).     As the model from Section~\ref{sec:analysis} individually fits three of the galaxies, we need to check whether this difference arises from these extra free parameters.  We therefore redo the analysis using only DM1 and DM2, and scaling all the cluster galaxies.  The resulting fit is worse than the original (rms$_s=0\farcs22$) but still significantly better than the fit without any smooth component.

The resulting mass map is shown in Figure~\ref{fig:need_DM} alongside our model from Section~\ref{sec:analysis}.  The mass map without a smooth component is very 'clumpy', whereas the latter is smooth.
\begin{figure}
\epsscale{1.0}
\plotone{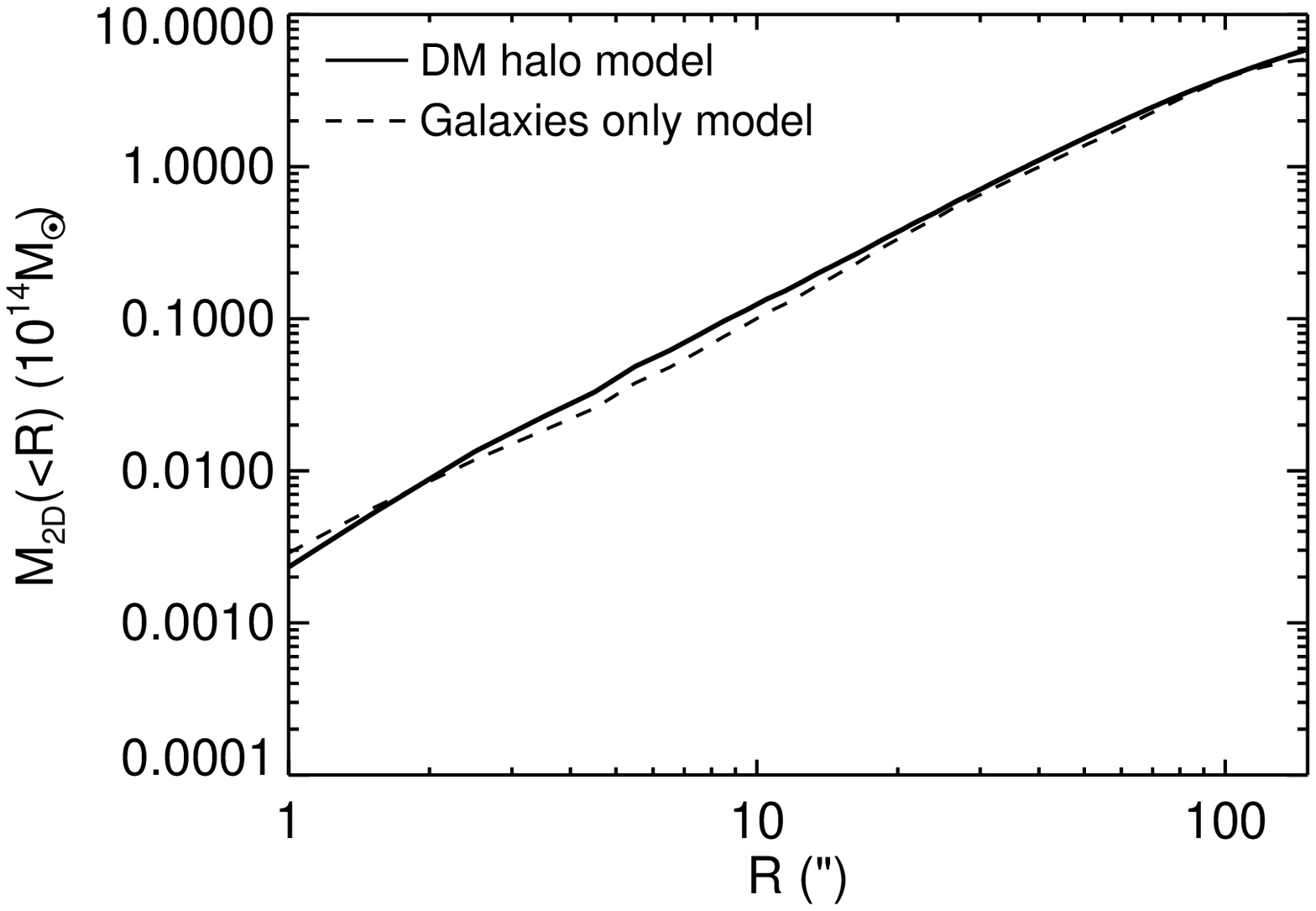}
\caption{The total mass as function of radius (centered on the BCG) for the two models.   Although the total mass within any given radii is similar, the pure-galaxy-mass map gives a significantly worse fit to the data (rms$_{s}=0.12$ vs rms$_{s}=1.62$), confirming the need for a large scale dark matter halo to accurately fit the data (see Section~\ref{sec:need_DM}). \label{fig:need_DM_total}  }
\end{figure}
As expected, the enclosed mass derived from each model is comparable for any radius where we have observational constraints (see Figure~\ref{fig:need_DM_total}).  Therefore, the poorness of the 'clumpy' fit is not due to the fact that it is not massive enough to reproduce the lensing constraints.  Moreover, it is worth noting that the 'clumpy' model is not very satisfactory in the sense that it describes the 
cluster galaxies as being very massive (around a few times $10^{12} M_\sun$ on average), which is not compatible with independent galaxy-galaxy lensing probes of
cluster galaxy masses \citep[see e.g., ][]{mandelbaum2005b}.  We therefore interpret the difference in the goodness of fit as evidence for the dark matter being distributed smoothly in the cluster, with only small perturbations from the cluster galaxies.  This is further supported by the X-ray emission (see Section~\ref{sec:bimodal}).

\begin{deluxetable*}{lllll} 
\tablecolumns{5} 
\tablewidth{0pc} 
\tablecaption{Goodness of fit for different cut off magnitudes}
\tablehead{ \colhead{Cut off magnitude in $K$}   & \colhead{\# of galaxy members} &  \colhead{$\log$~E} & \colhead{rms$_s$~($\arcsec$)} & \colhead{rms$_i$~($\arcsec$)}}
\startdata 
19.6 & 197	& $-205$ &	0.12 &	1.49 \\
19.0 & 145	& $-192$ &	0.12 &	1.05 \\
18.0 & 110	& $-238$ &	0.14 &	1.77 \\
17.0 &  62		& $-284$ &	0.17 &	1.87 \\
16.0 &  35		& $-216$ &	0.12 &	1.18 \\

\enddata 
\tablecomments{The $\chi^2$ and Evidence (E) for different cutoff magnitude in the galaxy catalogue.  All the models are optimised in the same way, the only difference being the number of galaxy members which are optimised using the scaling relations in section~\ref{sec:scaling}. }
\label{tab:gal_cutoff} 
\end{deluxetable*} 

\subsection{Sensitivity to the galaxy scale perturbers}
\label{sec:sens_gal}
We have already seen that individual galaxies are important to the overall lens model if they are close to a multiply imaged system (see also Section~\ref{sec:blind_test}), but in general the cluster galaxies only add small perturbations to the overall 
mass distribution of the cluster, with $\sim5$-$6$\% of the total mass being associated with the galaxy sized halos (excluding the BCG, see Figure~\ref{fig:m_1d}).  
To check whether those small scale perturbations are important, we try using a subset of the galaxy catalog to see if this affects the results.  
Our original catalog contains $197$ cluster members down to $K=19.6$ (including the individually fitted galaxies), but we also create catalogs using a cut off magnitude of 
$K=19, 18, 17, 16$ with $145, 110, 62, 35$ members respectively.  
We find that the overall quality of the fit is only weakly affected (see Table~\ref{tab:gal_cutoff}), suggesting that for the purposes of lensing studies, it is sufficient to include the brightest 
galaxy members in the modeling in addition to those which clearly locally perturb given multiply imaged systems.  
This point is important in order to save computing time, as increased number of clumps, even if modelled by scaling, significantly increases the required CPU (Central Processing Unit) time.

\begin{figure*}
\epsscale{.4}
\plotone{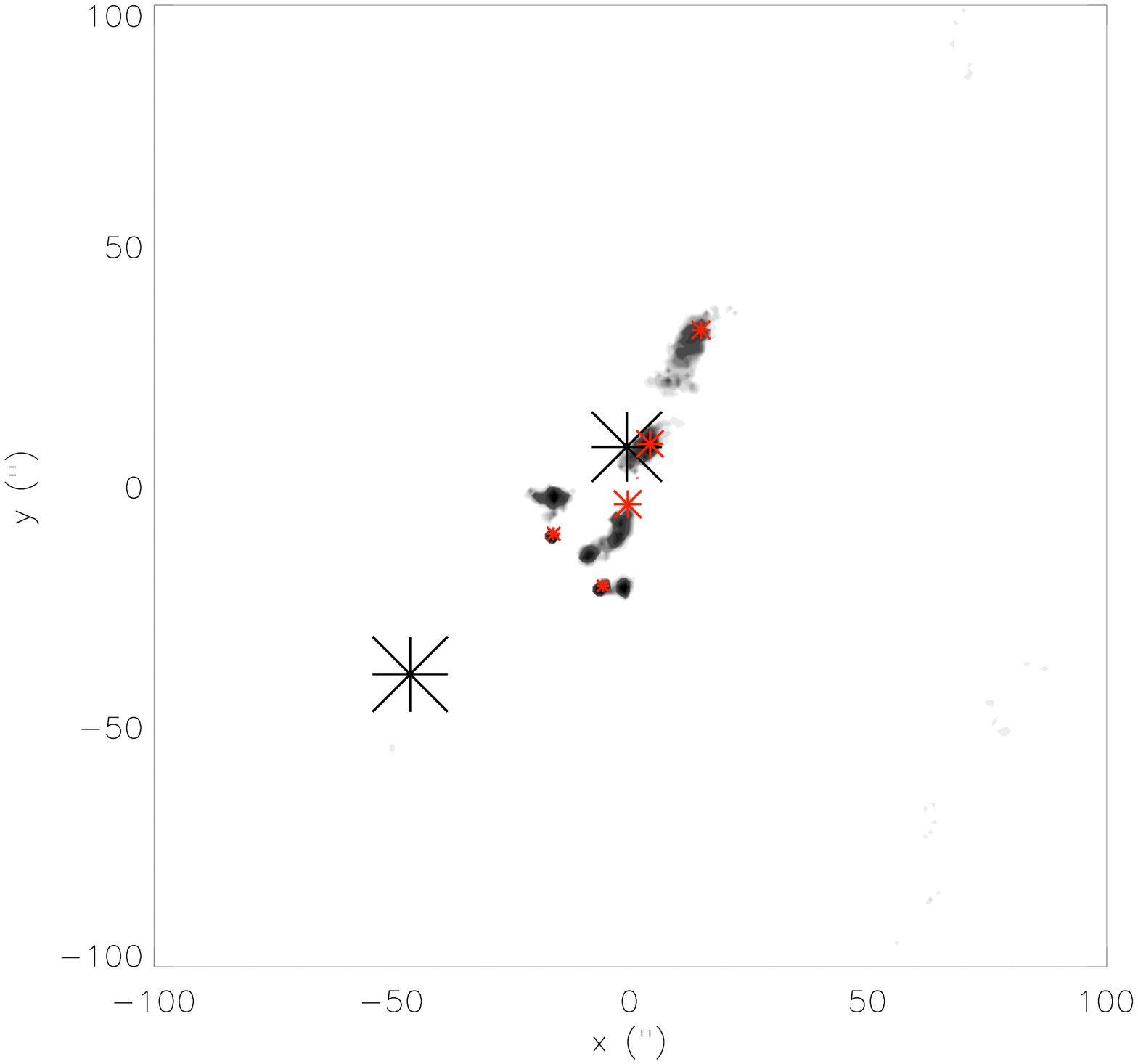}
\plotone{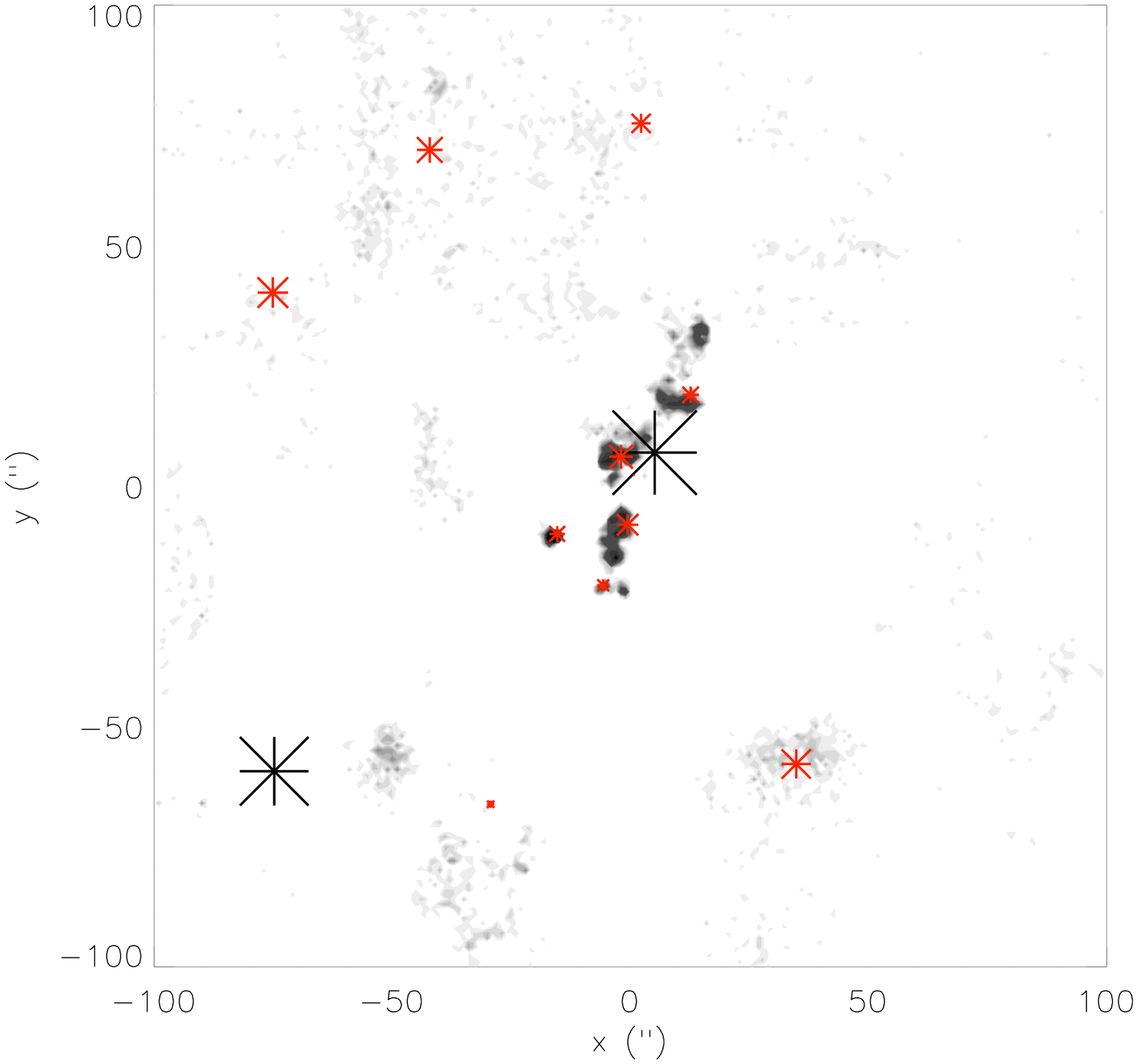}
\caption{Density plots for the positions of the galaxy sized clumps for the models discussed in section~\ref{sec:blind_test}.  The locations of the clumps for the best model are marked with red stars.  The size of the stars is proportional to their velocity dispersion.  The two big black stars show the two large dark matter clumps, their size again proportional to the velocity dispersion.  {\it Left panel:}  Configuration (A):  The model consists of five freely placed galaxy sized halos in addition to DM1 and DM2.   {\it Right panel:}  Configuration (B)  The model consists of ten freely placed galaxy sized halos in addition to DM1 and DM2.   Both configurations place galaxy scaled clumps where the BCG and the galaxies responsible for the local splitting of S1 are located.  The maps are $200\arcsec\times200\arcsec$ and are centered on the BCG with North being up and East being left. \label{fig:blind_test}}
\end{figure*}

\subsection{Blind tests - can we localize galaxy scale subtructure?}
\label{sec:blind_test}
Normally galaxy sized perturbers are added to the model by placing a dark matter clump at the location of a known cluster member.  This method is by construction not able to detect dark substructure directly.  To test how sensitive we are to these substructures, luminous or dark, we perform 'blind tests', where 
we model the cluster with (A) five and (B) ten freely placed galaxy sized dark matter clumps, in addition to the two large scale dark matter clumps DM1 and DM2.  

In order to limit the number of free parameters, we fix both $a=1$~kpc and $s=40$~kpc for the galaxy sized dark matter clumps, but we optimize all the parameters of the large scale clumps as before.   We allow the positions of the galaxy scale clumps to vary from $-100\arcsec$ to $100\arcsec$, corresponding roughly to the ACS field of view.  The velocity dispersion, $\sigma_{\mathrm{dPIE}}$, of each clump is allowed to vary from $0$-$500$~km~s$^{-1}$.

The results are shown in Figure~\ref{fig:blind_test}, showing a density plot of the position of the clumps for the 1000 best realizations (in the lowest $\chi^2$ sense), for both configurations (A) and (B).  The locations of the clumps corresponding to the best realization are marked by stars.
We find that both configurations give a good fit to the data (rms$_s=0.17$ and rms$_s=0.19$ for (A) and (B) respectively).  Configurations (A) and (B) are consistent with each other: setup (A) finds five well defined clumps. Setup (B) finds the same well defined five clumps, whereas it tries to marginalize the
five extra clumps that seem not to be needed in the optimization.

For both configurations, the model places two large halos where the BCG is, clearly demonstrating the need for significant mass at the location of the BCG galaxy.
Another common feature, is the lack of a galaxy sized clump in the vicinity of DM2, suggesting a very flat profile is preferred there.  Both configurations place clumps in the positions of galaxies \#1028 and \#993, which are responsible for the local splitting of S1.  This shows that lensing analysis can reliably detect galaxy sized dark substructure if it causes the local splitting of images, although the lensing map is not very sensitive to galaxy sized substructure in general (see also Section~\ref{sec:sens_gal}). 

\section{Bimodality of the Mass Distribution: Evidence of a Merger}
\label{sec:bimodal}
Our strong lensing analysis shows that Abell 2218 has a bimodal mass distribution: even if the dominant mass component is the BCG  and DM1 halos, the second large scale dark matter clump DM2 is also significant, contributing  around $20\%$ of the total mass within $100\arcsec$.
This second smooth mass component is associated with the bright cluster galaxy in the south-east that we call $\#617$.    To further interpret this bimodality, we look at two different available probes of the
cluster: the X-ray map and the velocity of cluster members.

\begin{description}
 \item[The X-ray map]
The X-ray flux map in the central parts of the cluster shows a complex morphology with no outstanding central peak. The offset between the X-ray peak at (R.A., Dec.)=$(248.967, 66.210)$
and the BCG is significant, $\sim 20\arcsec$, with the X-ray peak located in the direction of DM2, but there is no evident peak in the X-ray emission in the vicinity of DM2.  The X-ray flux map within about 1 arcmin of the peak of the X-ray emission is clearly elongated in the SE-NW direction, but becomes more spherical with increasing distance from the cluster center (see Figure~\ref{fig:massmap}).  Comparing the contours for the X-rays and the mass map we see that the elongation of the two are very similar, although the X-rays become more spherical at large radii.  To get more quantitative values for comparison of the two maps, we fitted a 2-dimensional $\beta$ model \citep{cavaliere1978} to both maps.

For the X-ray map, the best fit beta-model to the inner 2 arcmin by 2 arcmin (centered on the X-ray centroid) has an eccentricity ($\epsilon^\beta\equiv\sqrt{1-(B/A)^2}$) of $\epsilon^\beta_X(\mathrm{inner})=0.27\pm0.12$ and a position angle of $\theta_X(\mathrm{inner})=39\degr\pm16\degr$ (measured anti-clockwise from West). Fitting to the the 4 arcmin by 4 arcmin X-ray map (centered on the X-ray centroid) results in eccentricity of $\epsilon^\beta_X(\mathrm{outer})=0.18\pm0.08$ and a position angle of $\theta_X(\mathrm{outer})=17\degr\pm14\degr$.   A similar analysis for the mass map in a 5 arcmin by 5 arcmin (centered on the BCG) gives an eccentricity of the overall mass map as $\epsilon^\beta_{\mathrm{mass}}\approx0.25$ with $\theta_{\mathrm{mass}}\approx39\degr$, consistent with that found for the X-ray map, in particular in the inner regions.

\item[Distribution in velocity space]
If a merger has taken place in Abell~2218, then this should be imprinted
in the velocity distribution of the cluster members: we should be able to
identify two structures in velocity space; one associated with the BCG galaxy, and 
one associated with galaxy $\#617$.  

\citet{girardi1997} studied the structure of Abell 2218 using the spectroscopic data from \citet{leborgne1992}.  They found evidence for two structures (labelled MS1 and MS2)  separated by $2000$~km~s$^{-1}$.  The larger of these two structures, MS2, contains both the BCG and the brightest galaxy in DM2.  The two structures are superimposed along the line of sight and do not correspond exactly to the clumps found for the strong lens modeling.   Such superimposed structures are hard to separate using lensing as it is not sensitive to the 3D distribution of the matter.

To look for substructure associated with the BCG and $\#617$ we have calculated the separation ($\Delta v = c (z_i-z_j)/(1+z_i)$) between them and the other cluster members using spectroscopic data from \citet{leborgne1992}.  
Out of the 50 galaxies in the catalog we found 13 with $\Delta v < 500$~km~s$^{-1}$ and 25 with $\Delta v < 1000$~km~s$^{-1}$ for the brightest galaxy in DM2 while for DM1 the numbers were 11 and 21 respectively.  The $\Delta v<500$~km~s$^{-1}$ cut defines two structures without any common members, while the $\Delta v <1000$~km~s$^{-1}$  cut starts mixing the two groups.   
While all the galaxies we find with a $\Delta v < 500$~km~s$^{-1}$ for the BCG and $\# 617$ would belong to MS2 found by \citet{girardi1997}, the data suggests that MS2 may be further subdivided into two smaller structures, corresponding to the DM1 and DM2 found by the lens models.
\end{description}

The X-ray data and the distribution of the cluster members in velocity space, indicates that the bimodal mass distribution is caused by a merger.   The main cluster is the one associated with the BCG (DM1 and BCG in the lensing analysis) which the second cluster associated with $\#617$ (DM2 in the lensing analysis) has merged with, thus displacing the center of the X-ray peak from the BCG galaxy.  

\section{Conclusions}
\label{sec:conclusions}
We have reconstructed a mass map of the rich galaxy cluster Abell 2218 using strong lensing constraints.  Our model is based on 7 multiply imaged systems and 1 arc with spectroscopic redshifts,  and 6 systems without  spectroscopic redshifts, of which 5 are new candidate systems proposed in this work.  The model is sampled and optimized in the source plane by a Bayesian Monte Carlo Markov Chain implemented in the publicly available software Lenstool.  Our best model has rms$_s=0\farcs12$ in the source plane, corresponding to rms$_i=1\farcs49$ in the image plane.

We have found, in agreement with previous models of Abell 2218, that the mass distribution is bimodal.
We find that DM2 is larger and with a flatter core than previous models have found.  The flatness of the profile near DM2 is further supported by our 'blind tests', which do not place a galaxy sized component near its center.  The BCG and DM1 are the dominant component of the mass model in the inner regions  ($<100\arcsec$) of the cluster.  We have analyzed the distribution of galaxies in velocity space, finding evidence for two substructures, separated by $\sim 1000$~km~s$^{-1}$, corresponding to DM1 and DM2.  Although both the light and the X-ray contours are consistent with the mass map, the center of the X-ray emission is offset from the central peak of the BCG.  We find that the X-ray data and the distribution in velocity space support the interpretation that the bimodal mass distribution arises from a cluster merger.

We have explored the degeneracy of the mass model, both those inherent to the dPIE profile and those arising from the mass model components themselves.   For the dPIE, our results are in agreement with those of \citet{jullo2007}.   We find that the large scale dark matter clumps are a necessary component of the model, i.e., using only the dark matter halos associated with the galaxies does not give a good fit to the data (rms$_s=1\farcs62$ vs. rms$_s=0\farcs12$), even if they are allowed to become much more massive. 

At around $100\arcsec$ the two large scale halos contribute $\sim85\%$ of the enclosed projected mass, while the BCG contributes $\sim9\%$ and the remaining cluster galaxies $\sim6\%$.  We have performed 'blind tests' to check where the model requires a galaxy scale component to reproduce the lensing constraints, and find that both the BCG and galaxies which locally perturb given multiply imaged systems are reliably reclaimed.    However, we find that the inclusion of the cluster galaxies (excluding the BCG) only weakly affects the model unless they locally perturb a multiply imaged system.  Assuming that mass scales with light, this shows that strong lensing constraints can reliable detect substructure, dark or luminous, if the substructure is massive or locally perturbs a system.

The accurate mass map we have presented is made available to the community and can be used to exploit Abell 2218 as a gravitational telescope, probing the high redshift universe.  In this work we have fixed the cosmology, but with the increased number of constraints it may be possible to simultaneously fit it with the mass distribution.  However, the bimodal structure of Abell 2218, and the remaining uncertainty in the mass model (around $\sim1\arcsec$ in the image plane) may make it challenging to reach competitive accuracy in the derived cosmological parameters.

\acknowledgments

We thank S\'ebastien Bardeau and Genevi\`eve Soucail for giving us access to the weak lensing data.  \'A.E. thanks Bo Milvang-Jensen, C\'ecile Faure, Chlo\'e F\'eron, John McKean, Kim Nilsson and Paul Vreeswijk for helpful discussions relating to this work.
The Dark Cosmology Centre is funded by the Danish National Research Foundation.  This work was supported by the European Community's Sixth Framework Marie Curie Research Training Network Programme, Contract No. MRTN-CT-2004-505183 "ANGLES".  We thank the Danish Centre for Scientific Computing for granting the computer resources used.  This research was supported in part by the National Science Foundation under Grant No. PHY99-07949.  JR is grateful to the California Institute of Technology for its support.  JPK acknowledges support from CNRS.  KP acknowledges support from IDA. The authors recognize and acknowledge the very significant cultural
role and reverence that the summit of Mauna Kea has always had
within the indigenous Hawaiian community.  We are most fortunate
to have the opportunity to conduct observations from this mountain.




\appendix
\section{The Dual Pseudo Isothermal Elliptical Mass Distribution}
\label{app:piemd}
The Pseudo Isothermal Elliptical Mass Distribution (PIEMD) was first introduced by \citet{kassiola1993a} and it and its variants are frequently used in lensing analysis \citep[see e.g.,][]{kneib1996, keeton1998,smith2005, limousin2007, natarajan2007, richard2007}.  The main advantage of using the PIEMD for lensing analysis is that its potential, as well as its first and second partial derivatives, can be expressed analytically to derive deflections, distortions, amplifications and time delays for any ellipticity  (see also \citet{keeton1998}).  One of the most commonly used variants, and the one used in this work, is a two component PIEMD with both a core radius and a scale radius.  This profile is sometimes referred to as the "trunctuated PIEMD" or simply the "PIEMD", although it differs in significant ways from the original PIEMD proposed by \citet{kassiola1993a}.   An equivalent family of mass models was introduced independently by \citet{merritt1996} who used it as a starting point in an investigation of the orbital properties of triaxial stellar systems.  In this appendix we give a self-contained description of this profile, which we will call the 'dual Pseudo Isothermal Elliptical mass distribution' (dPIE) and present its derived quantities most relevant to lensing.

\subsection{The 3D density profile}
The 3D density distribution of the dPIE is 
\begin{equation}
\rho(r)={\rho_0 \over (1 + {r^2 / a^2}) (1 + {r^2 / s^2})}; \ \ \ \ s>a.
\end{equation}
This distribution represents a spherical system with scale radius $s$, core radius $a$ and central density $\rho_0$ \footnote{We note that $s$ corresponds to $r_{\mathrm{cut}}$ and $a$ to $r_{\mathrm{core}}$ in previous publications (e.g. \citet{limousin2007,jullo2007}), but we adopt this new notation to avoid confusion, as $s$ is not a 'cut-off'  nor a trunctuation radius, but the a scale radius, with $\rho\sim r^{-4}$ for $r>>s$ (see also Section~\ref{sec:piemd_mass} for how the mass depends on $s$).}. 
In the center $\rho\approx \rho_0/(1 + r^2 / a^2)$ which describes a core with central density $\rho_0$. The core is not strictly 
isothermal (in which case $\rho\sim(1 + r^2 / a^2)^{-3/2}$ 
\citep{binney1987}) but in the transition region, $a\la r\la s$, we have
$\rho\sim r^{-2}$ as for the isothermal sphere. In the outer parts the
density falls off as $\rho\sim r^{-4}$. 

The density distribution can be rewritten as
\begin{equation}
\rho(r)={\rho_0 \over (1 + (\eta\xi)^2) (1 + \xi^2)}; \ \ \ \ \eta > 1
\end{equation}
with $\xi\equiv r/s$ and $\eta\equiv s/a$. Thus $s$ acts as a scale parameter 
and $\eta$ is a shape parameter. 

\subsection{The 2D density profile}

The projected density, which is the relevant quantity for lensing, is given by
\begin{equation}
\Sigma(R)  =  2\int_R^\infty {\rho(r) r \over \sqrt {r^2-R^2}}\d r  = \Sigma_0 {as \over s-a}         \left ( {1 \over \sqrt{a^2+R^2}}- {1 \over \sqrt{s^2+R^2}} \right )
\end{equation}
with
\begin{equation}
\Sigma_0=\pi \rho_0 {as \over s+a}.
\end{equation}
and $R$ is the 2D radius.
For a vanishing core radius it reduces to
\beq
\Sigma(R)=\Sigma_0{a \over R} \left ( 1- {1 \over \sqrt{1 + s^2/R^2} } \right ); \ \ \ \ \ a << R
\eeq
which is the surface mass profile proposed by \citet{brainerd1996} for modeling galaxy-galaxy lensing. In this special case, $a$ becomes a 
simple amplitude scaling parameter and, as shown below, $s$ is the half-mass 
radius, $r_h$.

Defining the critical surface mass density 
\beq
\Sigma_{\rm crit}\equiv  {c^2\over 4\pi G} {D_{S} \over D_{L}D_{LS} }.
\eeq
where $D_L$, $D_S$, and $D_{LS}$ are the angular diameter
distances to the lens, the source, and between the lens and source
respectively,
the convergence is
\beq
\kappa (R) \equiv {\Sigma (R)\over \Sigma_{\rm crit}} 
\eeq
and the corresponding shear is
\beq
\gamma(R)={\Sigma_0 \over \Sigma_{\rm crit}}  {as \over s-a} \left [
2 \left ( {1 \over a + \sqrt{a^2+R^2}} - {1 \over s + \sqrt{s^2+R^2}} \right ) +
\left ( {1 \over \sqrt{a^2+R^2}}- {1 \over \sqrt{s^2+R^2}} \right ) \right ].
\eeq

\subsection{Mass relations}
\label{sec:piemd_mass}
The mass inside physical radius $r$ is
\begin{equation}
M_{3D}(r) =  4 \pi \int_0^r \rho(\tilde{r}) \tilde{r}^2  \d \tilde{r}  = 4 \pi \rho_0 {a^2s^2 \over s^2-a^2}        \left ( s \arctan \left ({r \over s} \right ) -              a \arctan\left ({r \over a} \right ) \right )
\end{equation}
and the mass inside projected radius $R$ is
\begin{equation}
\label{eq:mass2D}
M_{2D}(R)  =  2 \pi \int_0^R \Sigma(\tilde{R}) \tilde{R} \d \tilde{R}  = 2 \pi \Sigma_0 {as \over s-a} 
          \left ( \sqrt{a^2+R^2}-a - \sqrt{s^2+R^2}+s \right ).
\end{equation}
The dPIE has a finite total mass, given by 
\begin{equation}
M_{\rm TOT}=2 \pi^2 \rho_0 {a^2s^2 \over a+s}=2 \pi \Sigma_0 as.
\end{equation}

The half-mass radius defined by $M_{3D}(r_h)=M_{\rm TOT}/2$ obeys the
relation
\beq
s \arctan \left ({r_h \over s} \right ) - 
a \arctan \left ({r_h \over a} \right ) = {\pi \over 4}(s-a).
\eeq
For $a \ll r_h$, $s$ 
\beq
r_h \approx s+  {\pi \over 2} a
\eeq
i.e., $r_h \ge s$ with equality for $a \to 0$.  Similarly, the effective radius defined by $M_{2D}(R_e)=M_{\rm TOT}/2$ is
\beq
R_e=\frac{3}{4} \sqrt{a^2+\frac{10}{3} a s+ s^2}
\eeq
which for $\eta^{-1}\ll 1$ ($a \ll s$) reduces to
\beq
R_e\approx{3 \over 4} s + {5 \over 4} a, 
\eeq
i.e., $R_e\approx(3/4) r_h$ for $a \to 0$. 

\subsection{The potential}
The gravitational potential $\Psi=-\Phi$ is
\begin{equation}
\Psi(r)  =  G \int_r^\infty {M_{3D}(r) \over r^2}  \d r =  4 \pi G \rho_0 {a^2s^2 \over s^2-a^2} 
          \left ( 
{s \over r} \arctan \left ({r \over s} \right ) -
{a \over r} \arctan \left ({r \over a} \right ) + 
          {1 \over 2} \ln \left ( s^2 + r^2 \over a^2 + r^2 \right ) \right )
\end{equation}
and the central potential is
\beq
\Psi(0) = 4 \pi G \rho_0 {a^2s^2 \over s^2-a^2} \ln \frac{s}{a} 
\eeq
which is finite for finite $a$.

The projected potential is given by
\begin{eqnarray}
\Psi(R) & = & 2 G \int_R^{R_L} {M_{2D}(\tilde{R}) \over \tilde{R}}  \d \tilde{R} \nonumber \\
      & = & 4 \pi G \Sigma_0   {as \over s-a} 
          \left ( \sqrt{s^2+R^2}-\sqrt{a^2+R^2}+
          a \ln ( a + \sqrt{a^2+R^2} ) - s \ln ( s + \sqrt{s^2+R^2} ) \right )  \nonumber \\
          & &   +\  {\rm constant,}
\end{eqnarray}
where $R_L$ is a limiting radius leading to the constant term. The deflection 
then is
\begin{equation}
\alpha(R)  =  - {2\over c^2}{ D_{LS} \over D_S} {\d \Psi \over \d R}  = {8 \pi G \over c^2} {D_{LS} \over D_S} \Sigma_0 {as \over s-a} f(R/a,R/s)
\end{equation}
where
\begin{equation}
f(R/a,R/s)\equiv   \left ( {R/a \over 1 + \sqrt{1^2+(R/a)^2}}-{R/s \over 1 + \sqrt{1^2+(R/s)^2}} 	  \right ).
\end{equation}

\subsection{Ellipticity}
An elliptical projected mass distribution with ellipticity
$\epsilon\equiv (A-B)/(A+B)$, where $A$ and $B$ are the semi major and minor 
axes respectively, can be introduced by substituting $R\to \tilde R$, with 
\beq
\tilde{R}^2={X^2 \over (1+\epsilon)^2} + {Y^2 \over (1-\epsilon)^2},
\eeq
where $X$ and $Y$ are the spatial coordinates along the major and minor 
projected axes, respectively. With this definition of the ellipticity, the 
quantity $e\equiv 1-B/A$ is related to $\epsilon$ through 
$e=2\epsilon/(1+\epsilon)$.  For further details we refer to \citet{kassiola1993a}, and note that in their notation the model discussed in this paper corresponds to
\beq
J_{as,3/2}^*={as \over s-a} (a^{-1}I^*_{a,1/2}-s^{-1}I^*_{s,1/2}).
\eeq
and the analytical potential is given by solving
\begin{eqnarray*}
\frac{\delta^2\Phi}{\delta X^2}=\Re \frac{\delta J_{as,3/2}^*}{\delta X}  ;\hspace{1cm} & \frac{\delta^2\Phi}{\delta Y^2}=\Im \frac{\delta J_{as,3/2}^*}{\delta Y}  ;\hspace{1cm}& \frac{\delta^2\Phi}{\delta X \delta Y}=\Im \frac{\delta J_{as,3/2}^*}{\delta X}=\Re \frac{\delta J_{as,3/2}^*}{\delta Y}.
\end{eqnarray*}

\subsection{The dPIE in Lenstool}
The dPIE is incorporated into the Lenstool software using the full elliptical expressions.  The profile is given by specifying 8 parameters, its redshift $z$, its central position (R.A.,Dec.), its ellipticity and orientation $(\hat{\epsilon}, \theta_{\hat{\epsilon}})$, the core radius $a$, the scale radius $s$ and a fiducial velocity dispersion $\sigma_{dPIE}$.
This fiducial velocity dispersion is related to the deflection, $\alpha$, by
\begin{equation}
\alpha=\frac{a+s}{s} E_0 f(R/a,R/s)
\end{equation}
 where 
 \begin{equation}
E_0=6\pi \frac{D_{LS}}{D_S}\frac{\sigma_{dPIE}^2}{c^2}.
\end{equation}
In terms of $\Sigma_0$, $a$ and $s$ (or $\rho_0$, $a$ and $s$) $\sigma_{\mathrm{dPIE}}$ is given by
\begin{equation}
\sigma_{\mathrm{dPIE}}^2=\frac{4}{3} G \Sigma_0 \frac{a s^2}{s^2-a^2}=\frac{4}{3} G \pi \rho_0  \frac{a^2 s^3}{(s-a)(s+a)^2}.
\end{equation}

The ellipticity is defined by
\begin{equation}
\hat{\epsilon}=\frac{A^2-B^2}{A^2+B^2}
\end{equation}
corresponding to 
\begin{equation}
e=1-\frac{B}{A} = 1-\frac{\sqrt{1-\hat{\epsilon}}}{\sqrt{1+\hat{\epsilon}}}.
\end{equation}
Its direction,  $\theta_{\hat{\epsilon}}$, is measured anti-clockwise from the West and is related to the position angle (P.A.) by P.A.$=\theta_{\hat{\epsilon}}-90\degr$.

\subsection{Velocity dispersions}
\label{sec:app_veldisp}
Although $\sigma_{dPIE}$ is a fiducial velocity dispersion, we wish to relate it to the measured velocity dispersion of galaxies, assuming that their profile is described by a dPIE.  In this derivation we will look at the most simple case, assuming no anisotropy and a spherically symmetric dPIE profile.  Even under these assumptions, the central velocity dispersion and the velocity dispersion profiles for the dPIE cannot 
be expressed analytically and must be computed numerically. The 
velocity dispersion \citep[][]{binney1987} is
\beq
\sigma^2(r)={G \over \rho(r)} \int_r^\infty {M_{3D}(r) \rho(r) \over r^2} \d r
\eeq
and the 
projected (line-of-sight) velocity dispersion is
\beq
\sigma_P^2(R)={2 G \over \Sigma(R)} \int_R^\infty {M_{3D}(r) \rho(r) \over r^2} 
\sqrt{r^2-R^2} \d r .
\eeq
We can also compute the average line-of-sight velocity dispersion inside
a given radius $R$ as
\begin{equation}
\label{eq:avsigma}
\left<\sigma_P^2\right>(R)= {2 \pi \int_0^R \sigma_P^2(R) 
\Sigma(R) R \d R \over M_{2D}(R)}.
\end{equation}
This is the quantity measured by velocity dispersion measurements, and for galaxies the relevant radius corresponds roughly to the slit width used for the spectrograph.  For the galaxies in our model (see Table~\ref{tab:model}) and a spectrograph resolution of  $\sim1\arcsec$, this corresponds to an observable region of $0.01 \la (R/R_e) \la 2$.  
\begin{figure}
\epsscale{.5}
\plotone{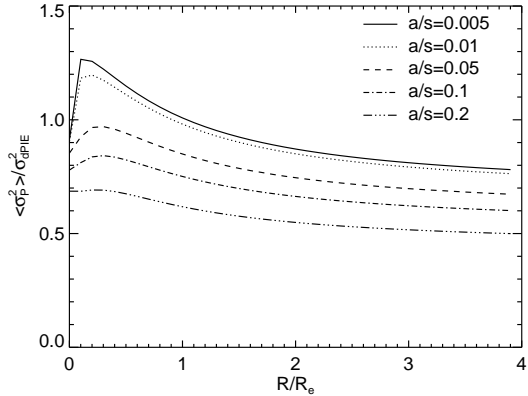}
\caption{ The squared measured velocity dispersion ($\left<\sigma_P^2\right>$) vs. the squared fiducial $\sigma_{\mathrm{dPIE}}^2$.   The ratio varies depending on both the observed radius $R$, and the ratio of $a$ and $s$.  For values similar to those found for the galaxies in the model (Table~\ref{tab:model}) we find $\sqrt{\left<\sigma_P^2\right>}\approx (0.85\pm0.10) \sigma_{\mathrm{dPIE}}$. \label{fig:piemd_vel_disp} }
\end{figure}
In Figure~A1 we plot the average line-of-sight velocity dispersion over the fiducial velocity dispersion $\sigma_{\mathrm{dPIE}}$ for different values of $\eta$.  We find that for $a/s$ values similar to those found in the modelling, $\sqrt{\left<\sigma_P^2\right>}/\sigma_{\mathrm{dPIE}}\approx 0.85\pm0.10$ in the observable region.

Finally, to facilitate comparison with velocity dispersion profiles from different profiles, we define a 'standard' velocity dispersion $\sigma_e$.  This velocity dispersion is  motivated by the velocity dispersion of the isothermal profile (for which $\sigma^2_{\mathrm{iso}}= 2 G M_{\mathrm{2D}}(R)/R$ for all $R$), calculated at the effective radius $R_e$:
\begin{equation}
\sigma^2_{e}\equiv 2 G \left<\Sigma\right>_e R_e
\end{equation}
where $\left<\Sigma\right>_e$ is the mean surface density at $R_e$.
For the dPIE profile, we find that:
\begin{equation}
\frac{\sigma^2_{e}}{\sigma_{\mathrm{dPIE}}^2} = \frac{2 (s^2-a^2)}{s \sqrt{ a^2+(10/3) a s +  s^2}}
\end{equation}
which goes to $2$ when the core radius $a$ goes to zero.  For $a/s$ values similar to those found in the modelling $\sigma_e/\sigma_{\mathrm{dPIE}}\approx1.3\pm0.2$.




\end{document}